\documentclass[
    prl, twocolumn, superscriptaddress, nofootinbib, amsmath, amssymb,
    aps, floatfix
]{revtex4-2}

\usepackage{graphicx,xcolor,setspace}
\usepackage[
    colorlinks=true,
    linkcolor=blue,
    urlcolor=blue,
    citecolor=blue
]{hyperref}

\usepackage{bm}
\newcommand{\bb}[1]{\bm{\mathrm{#1}}}
\newcommand{\du}{\mathrm d}
\newcommand{\dd}{\,\du}
\renewcommand{\Im}{\operatorname{Im}}
\renewcommand{\Re}{\operatorname{Re}}
\usepackage[capitalise]{cleveref}
\usepackage{siunitx}
\DeclareSIUnit{\year}{yr}
\newcommand\meV{\milli\electronvolt}
\newcommand\eV{\electronvolt}
\newcommand\keV{\kilo\electronvolt}
\newcommand\kms{\kilo\meter/\second}
\usepackage{mhchem}
\newcommand{\URuSi}{\ce{URu_2Si_2}}

\newcommand{\rhat}{\hat{\bb{r}}}

\begin{document}

\title{Determining Dark Matter--Electron Scattering Rates from the Dielectric Function}

\author{Yonit Hochberg}
\affiliation{Racah Institute of Physics, Hebrew University of Jerusalem, Jerusalem 91904, Israel}
\author{Yonatan Kahn}
\affiliation{Department of Physics, University of Illinois at Urbana-Champaign, Urbana, IL 61801, USA}
\affiliation{Illinois Center for Advanced Studies of the Universe, University of Illinois at Urbana-Champaign, Urbana, IL 61801, USA}
\author{Noah Kurinsky}
\affiliation{Fermi National Accelerator Laboratory, Batavia, IL 60510, USA}
\affiliation{Kavli Institute for Cosmological Physics, University of Chicago, Chicago, Illinois 60637, USA}
\author{Benjamin V. Lehmann}
\affiliation{Department of Physics, University of California Santa Cruz, Santa Cruz, CA 95064, USA}
\affiliation{Santa Cruz Institute for Particle Physics, Santa Cruz, CA 95064, USA}
\author{To Chin Yu}
\affiliation{Department of Physics, Stanford University, Stanford, CA 94305, USA}
\affiliation{SLAC National Accelerator Laboratory, 2575 Sand Hill Road, Menlo Park, CA 94025, USA}
\author{Karl K. Berggren}
\affiliation{Department of Electrical Engineering and Computer Science, Massachusetts Institute of Technology, Cambridge, MA 02139, USA}
\date\today

\begin{abstract}
We show that the rate for dark matter--electron scattering in an arbitrary material is determined by an experimentally measurable quantity, the complex dielectric function, for any dark matter interaction that couples to electron density. This formulation automatically includes many-body effects, eliminates all systematic theoretical uncertainties on the electronic wavefunctions, and allows a direct calibration of the spectrum by electromagnetic probes such as infrared spectroscopy, X-ray scattering, and electron energy-loss spectroscopy (EELS). Our formalism applies for several common benchmark models, including spin-independent interactions through scalar and vector mediators of arbitrary mass. We discuss the consequences for standard semiconductor and superconductor targets, and find that the true reach of superconductor detectors for light mediators exceeds previous estimates by several orders of magnitude, with further enhancements possible due to the low-energy tail of the plasmon. Using a heavy-fermion superconductor as an example, we show how our formulation allows a rapid and systematic investigation of novel electron scattering targets.

\end{abstract}

\maketitle

Dark matter (DM)--electron scattering was first proposed for sub-GeV DM detection less than a decade ago \cite{Essig:2011nj}, and there has been enormous theoretical \cite{Graham:2012su,Essig:2015cda,Lee:2015qva,Hochberg:2015pha,Hochberg:2015fth,Alexander:2016aln,Derenzo:2016fse,Hochberg:2016ntt,Kavanagh:2016pyr,Emken:2017erx,Emken:2017qmp,Battaglieri:2017aum,Essig:2017kqs,Cavoto:2017otc,Hochberg:2017wce,Essig:2018tss,Emken:2018run,Ema:2018bih,Geilhufe:2018gry,Baxter:2019pnz,Essig:2019xkx,Emken:2019tni,Hochberg:2019cyy,Trickle:2019nya,Griffin:2019mvc,Coskuner:2019odd,Geilhufe:2019ndy,Catena:2019gfa,Blanco:2019lrf,Kurinsky:2019pgb,Kurinsky:2020dpb,Griffin:2020lgd,Radick:2020qip,Gelmini:2020xir,Trickle:2020oki,Du:2020ldo} and experimental \cite{Essig:2012yx,Tiffenberg:2017aac,Romani:2017iwi,Crisler:2018gci,Agnese:2018col,Agnes:2018oej,Settimo:2018qcm,Akerib:2018hck,Abramoff:2019dfb,Aguilar-Arevalo:2019wdi,Aprile:2019xxb,Barak:2020fql,Arnaud:2020svb,Amaral:2020ryn} progress since then. Since electrons are not free particles, but are bound in atoms or delocalized across solids, they have favorable kinematics for light DM scattering. However, the rich complexity of condensed matter systems complicates the calculation of scattering rates. Not only do bound electrons have different wavefunctions than their free-particle counterparts \cite{AshcroftMermin}, many condensed matter systems exhibit collective electronic modes such as plasmons \cite{RevModPhys.28.184}. A formalism describing DM scattering with a single electronic state \cite{Essig:2015cda,Trickle:2019nya} can potentially miss important electron interaction and correlation effects, and must carefully account for `screening' where the electron density rearranges itself to partially cancel out DM-induced perturbations~\cite{Hochberg:2015fth}.

In this Letter we propose to bypass the single-particle formulation entirely, and frame the problem of DM--electron scattering in terms of matrix elements of the many-body electron density operator. This perspective is inspired by a classic paper on collective energy loss in solids \cite{nozieres1959electron}, and since it does not rely on a particular choice of eigenstates, it is equally applicable to \emph{all} systems: atoms, molecules, metals, insulators, or more exotic materials. Moreover, it intrinsically accounts for all electron interactions and correlations in the target by relating the scattering rate to an experimentally-measurable quantity, the complex dielectric function $\epsilon(\bb q, \omega)$. Crucially, since $\epsilon(\bb q, \omega)$ is defined as a linear response function, the response of the target to a momentum transfer $\bb q$ and energy deposit $\omega$ is determined by density matrix elements which are the \emph{same} whether measured by DM--electron scattering or by an electromagnetic probe \cite{raether2006excitation,schulke2007electron}. The assumption of linear response applies as long as DM interactions are weaker than electromagnetism.

The key result of this Letter is that the total scattering rate for DM with mass $m_\chi$ and velocity $\bb v_\chi$ in an arbitrary target is given by
\begin{equation}
    \label{eq:THEANSWER}
    \Gamma(\bb v_\chi) = \int\frac{\du^3\bb q}{(2\pi)^3}\,
        |V(\bb q)|^2 \, \left[2 \frac{q^2}{e^2} \,
        \Im\left(-\frac{1}{\epsilon(\bb q, \omega_{\bb q})}\right)\right]
    ,
\end{equation}
where $\omega_{\bb q} = \bb q \cdot \bb v_\chi - \frac{q^2}{2m_\chi}$, $q = |\bb q|$, $e$ is the electron charge, and $V(\bb q)$ is the non-relativistic DM-electron potential. The full derivation can be found in the Supplemental Material (SM), and follows mainly from the arguments made in Ref.~\cite{nozieres1959electron}. The target-dependent object which appears in the integrand,
\begin{equation}
    \label{eq:Wdef}
    \mathcal{W}(\bb q, \omega) \equiv
    \Im\left(-\frac{1}{\epsilon(\bb q, \omega)}\right) =
    \frac{\Im [ \epsilon(\bb q,\omega) ] }{|\epsilon(\bb q,\omega)|^2}
    ,
\end{equation}
is known as the loss function. The \emph{only} assumptions we have made about the DM interactions in deriving \cref{eq:THEANSWER} are (\textit{i}) that the non-relativistic Hamiltonian coupling DM to electrons takes the form $\hat{H}_{\mathrm{int}} = \sum_{i} V(\rhat_\chi - \rhat_i)$, depending only on the electron position operators $\rhat_i$ and no other operators such as spin or momentum, and (\textit{ii})~that $\hat{H}_{\mathrm{int}}$ can be treated perturbatively.

\begin{figure}[t]
    \centering
    \includegraphics[width=0.49\textwidth]{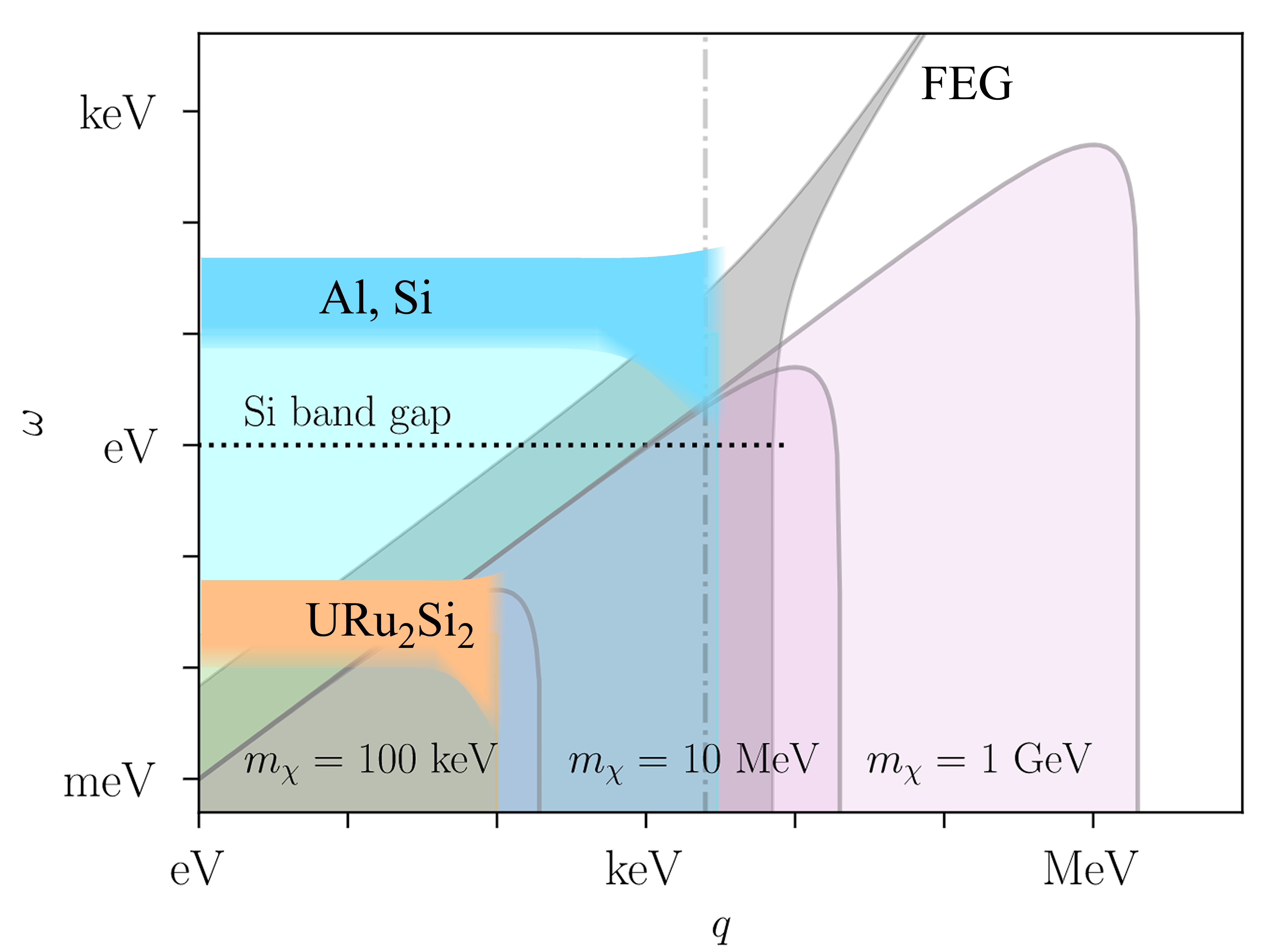}
    \caption{Schematic depiction of the relevant kinematics for sub-GeV DM. The shaded purple parabolas represent the kinematically-allowed region of $q$ and $\omega$ for the labeled DM masses, as in Ref.~\cite{Trickle:2019nya}, for a fixed DM speed $v_\chi = 10^{-3}$, with upper boundary $\omega = q v_\chi$ independent of $m_\chi$. The blue and orange shaded regions represent the support of the plasmon part of the loss function. The tail extends into the DM region for conventional materials such as Al and Si, and for heavy-fermion materials such as \URuSi, the plasmon peak lies in the DM region. The range of support for the free electron gas (FEG) loss function is shown in shaded gray, and can be used to approximate the rate in both superconductors and semiconductors over a limited range of $\omega$. The dot-dashed vertical line indicates the size of the Brillouin zone ($q \approx \SI{2.3}{\keV}$) of Si, while the horizontal dashed line indicates the band gap above which electron scattering can produce ionization.}
    \label{fig:kinematics}
\end{figure}

The consequences of \cref{eq:THEANSWER} are of immediate importance for DM--electron scattering.
Spin-independent Hamiltonians arise in many common benchmark models, including those for scattering through scalar and vector mediators. The presence of $\mathcal W$ implies that \emph{all} of these interactions are screened. The importance of screening was first noted for a kinetically-mixed dark photon mediator in a solid-state target~\cite{Hochberg:2015fth}, and later for a scalar mediator~\cite{Gelmini:2020xir}. Our results show that a scalar force which couples equally and oppositely to electrons and protons, whether short- or long-ranged, is screened \emph{exactly} like a kinetically-mixed dark photon. Furthermore, as long as ion contributions to the loss function are negligible (as in semiconductors well above the gap), forces that couple differently to nucleons and electrons are still screened identically. All such screening effects are invisible in a single-particle picture.

Since $\mathcal{W}(\bb q, \omega)$ is directly measurable through electromagnetic scattering, DM--electron scattering experiments can be calibrated experimentally, exactly as was done for DM absorption \cite{Hochberg:2016ajh,Hochberg:2016sqx,Bloch:2016sjj} using the measured real conductivity $\sigma_1(\omega) = (\omega/4\pi) |\epsilon(0,\omega)|^2 \mathcal{W}(0,\omega)$. The advantage of our approach is that the loss function can also be modeled semi-analytically in certain relevant energy and momentum regimes, and such models can be compared directly to data. This enables rapid assessment of candidate experimental targets, and potentially bypasses the need for numerical electron wavefunctions to determine the reach of novel detector materials. As shown in the SM, the loss function contains a sum over all possible final states of the target, and thus \cref{eq:THEANSWER} represents the maximum possible scattering rate which could be observed at any experiment sensitive to a particular subset of excitations, for example, electron--hole pairs.

In the following sections, we show that in a material with free carriers, the loss function scales as $\mathcal{W}(\bb q, \omega) \propto q$ at small $\omega$, which can be interpreted as the familiar screening which partially suppresses the $1/q^4$ enhancement characteristic of a light mediator. We then show that if $\mathcal{W}(0, \omega)$ is nonvanishing, a rate enhancement at small $q$ remains whenever $\omega$ is kinematically accessible. This behavior of the loss function can arise in two qualitatively different ways: interband transitions in insulators, and long-range plasmons which are generically present in all materials. As we will show, the low-energy plasmon tail may improve the sensitivity of superconducting detectors to light DM by several orders of magnitude, and materials with Fermi velocities \emph{slower} than $v_\chi$ may allow DM to access the bulk of the loss function rather than the tail. We illustrate these kinematic regimes in \cref{fig:kinematics}.

In this Letter, we adopt a generic form for the potential, $V(\bb q) = V(q) = \frac{g_\chi g_e}{q^2 + m^2_{\phi,V}}$, which is valid for DM coupling through a scalar mediator $\phi$ or vector $V$. We compute scattering rates by integrating \cref{eq:THEANSWER} over the DM velocity distribution, for which we take the Standard Halo Model (see SM for details). We frame our results in terms of a reference cross section $\overline\sigma_{e} = (\mu_{e\chi}^2/\pi)|V(q_0)|^2$ where $\mu_{e\chi}$ is the electron--DM reduced mass and $q_0 = \alpha m_e \simeq 3.7 \ {\rm keV}$ is a reference momentum. We show results for a light mediator $m_{\phi, V}^2 \ll q^2$, with heavy mediator results given in the SM (Fig.~\ref{fig:heavy-reach}). 

\section{Conventional superconductors}

\begin{figure}[t]
    \centering
    \includegraphics[width=\columnwidth]{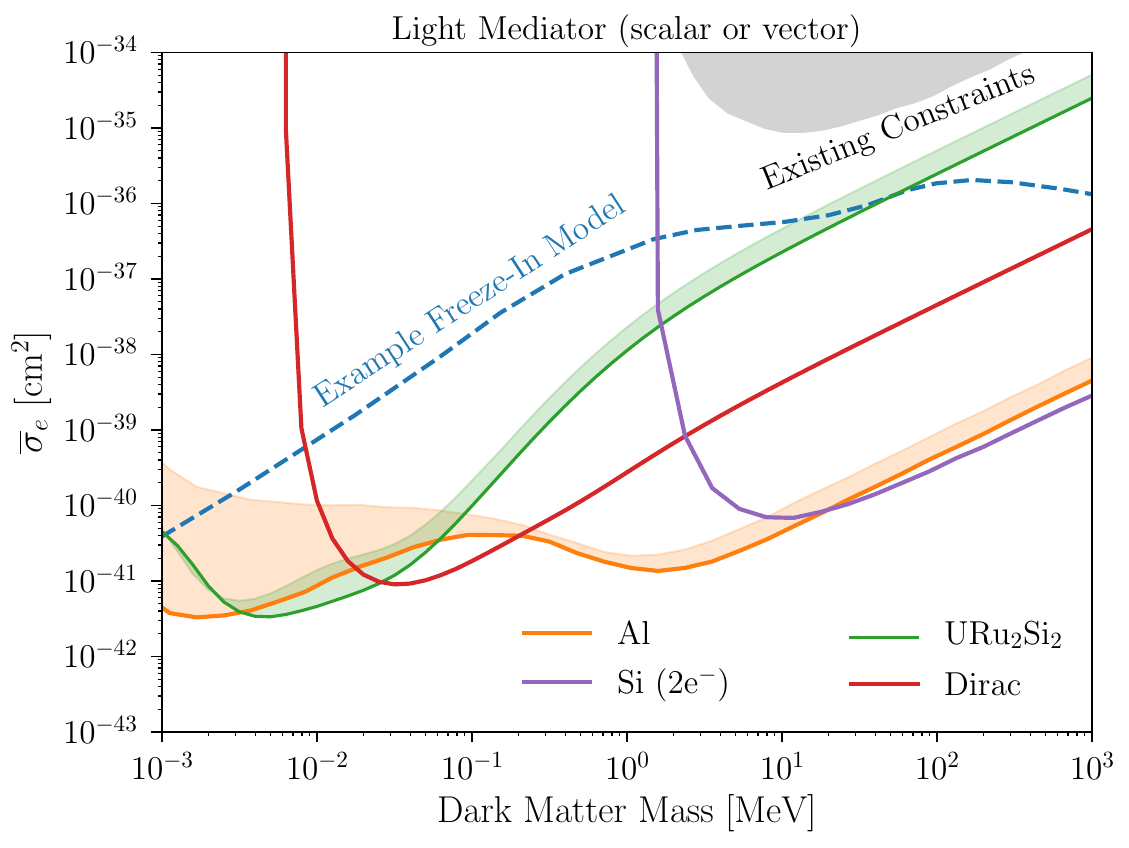}
    \caption{The projected 3-event reach of a 1~kg-yr exposure target of Al (orange), Si (purple), and \URuSi\ (green), computed for a light scalar or vector mediator using \cref{eq:THEANSWER}.
    For Al, the solid line uses $\mathcal{W}$ from Ref.~\cite{gibbons1976line}, and the top of the shaded region uses the FEG model, both with $\omega \in [\SI{1}{\meV}, \SI{1}{\eV}]$. The \URuSi~loss function is taken from Ref.~\cite{bachar2016detailed} with $\omega \in [\SI{1}{\meV}, \SI{74}{\meV}]$, and the shaded region spans $\mathcal{W}$ measured along two crystal axes. Si is treated as a FEG with a $2e^-$ threshold, using the ionization model of Ref.~\cite{Essig:2015cda}. We also show the reach for a Dirac material with density $\SI{10}{\gram/\centi\meter^3}$, gap $2\Delta = \SI{20}{\meV}$, Fermi velocity $v_F = 4 \times 10^{-4}$, background dielectric constant $\kappa = 40$, and Dirac band cutoff $\omega_{\mathrm{max}} = \SI{0.5}{\eV}$ (red); existing constraints from SENSEI~\cite{Barak:2020fql}, SuperCDMS HVeV~\cite{Amaral:2020ryn}, DAMIC~\cite{Aguilar-Arevalo:2019wdi}, Xenon10~\cite{Essig:2017kqs}, DarkSide-50~\cite{Agnes:2018oej}, and Xenon1T~\cite{Aprile:2019xxb} (shaded gray); and the theory target of a freeze-in model when the mediator is a kinetically-mixed dark photon \cite{Hall:2009bx,Essig:2011nj,Chu:2011be,Dvorkin_2019} (dashed blue). The corresponding plot for a heavy mediator is shown in the SM (Fig.~\ref{fig:heavy-reach}). 
    }
    \label{fig:reach}
\end{figure}

Ref.~\cite{Hochberg:2015pha} first proposed using superconducting metals such as aluminum (Al) as targets for DM--electron scattering. Ref.~\cite{Hochberg:2015fth} soon pointed out that long-range Coulomb forces among electrons would screen DM interactions if mediated by a kinetically-mixed dark photon. This effect was incorporated by multiplying the free-particle matrix element by $1/|\epsilon_{\mathrm{RPA}}(\bb q,\omega)|^2$, where $\epsilon_{\mathrm{RPA}}$ is the dielectric function of a free electron gas (FEG) in the random phase approximation (RPA) at zero temperature.

Even within RPA, our formalism identifies two important corrections to the DM interaction rate from Ref.~\cite{Hochberg:2015fth}. First, \emph{all} interactions coupling to electron density are screened, including a light scalar mediator and a non-kinetically-mixed vector mediator. This unifies the reach for all models considered in Ref.~\cite{Hochberg:2015fth}. Second, the analytic structure of the loss function imposed by causality implies a particular choice of branch cut in $\epsilon_{\mathrm{RPA}}$ differing from that used in  Ref.~\cite{Hochberg:2015fth} (see SM for details). 

The latter correction improves the projected sensitivity of conventional superconductor detectors to DM scattering through a light mediator by several orders of magnitude at low masses. We can understand this by examining $\epsilon_{\mathrm{RPA}}$ in the kinematic regime $q \ll k_F$, $\omega \ll q v_F$ relevant for sub-MeV DM scattering near the Fermi surface, where $k_F$ is the Fermi momentum and $v_F$ is the Fermi velocity, respectively \SI{3.5}{\keV} and $6.8\times10^{-3}$ in Al. The result is~\cite{dressel2002electrodynamics}
\begin{equation}
    \epsilon_{\mathrm{RPA}}(\bb q, \omega) \approx
        \frac{\lambda_{\mathrm{TF}}^2}{2 q^2} +
        i \frac{3 \pi \omega_p^2 \omega}{2 q^3 v_F^3}
    ,
\end{equation}
where $\lambda_{\mathrm{TF}} \simeq \SI{3.8}{\keV}$ is the Thomas-Fermi screening length and $\omega_p \simeq \SI{15}{\eV}$ is the plasma frequency. The imaginary part is typically smaller than the real part, so $\mathcal{W}(\bb q, \omega)$ scales as $\frac{\omega/q^3}{1/q^4} \sim \omega q$, a much softer screening than the $q^4$ implied from $1/|\epsilon|^2$.

Moving beyond RPA, we use the results of Ref.~\cite{gibbons1976line}, which fits to data a model containing both a 1-loop `local field' correction to the electron vertex and a $q$-dependent plasmon width $\Gamma_p/\omega_p \simeq 0.1\mbox{--}0.3$.  The fit implies that the contribution from the ion polarizability in Al is small, justifying our approximation that only electrons contribute to the loss function. The projected reach for a \SI{1}{\meV} threshold is shown in \cref{fig:reach} for a light mediator, with comparisons to previous results given in Fig.~\ref{fig:superconductor-reach} of the SM. The orange band reflects theoretical uncertainty in the proper form of the loss function in the energy range of interest~(see~SM). 

In most materials, the loss function features a plasmon with a Lorentzian lineshape peaked at $\omega_p$ \cite{RevModPhys.28.184,raether2006excitation} and a low-energy tail (see \cref{fig:kinematics} and SM). In the parametrization of Ref.~\cite{gibbons1976line}, $\mathcal{W}(q = 0, \omega)$ scales linearly with $\omega$ for $\omega \ll \omega_p$, and the plasmon tail dominates over the RPA contribution. Our results suggest that a kg-yr exposure of an Al target with a \SI{1}{\meV} threshold is sufficient to cover the entire freeze-in thermal relic target \cite{Hall:2009bx,Essig:2011nj,Chu:2011be,Dvorkin_2019} above \SI{10}{\keV}. However, this depends on the extrapolation of the plasmon tail to meV energies, and existing measurements only characterize the loss function at $\omega\gtrsim\SI{100}{\milli\eV}$ \cite{sun2016calculations}. Thus, additional measurements of $\mathcal{W}$ are crucial to accurately determine the sensitivity. There may also be contributions to $\mathcal{W}$ from coherent scattering with the Cooper pair condensate for energies $\omega \simeq 2\Delta$, as well as finite-temperature effects. We leave investigation of these effects for future work~\cite{future}.

\section{Semiconductors}

In a typical semiconductor like silicon (Si) with a gap $E_g \sim \SI{}{\eV}$, an energy deposit $\omega  \simeq E_g$ requires a momentum deposit $q \geq E_g/v_\chi \sim \SI{}{\keV}$ for $v_\chi \sim 10^{-3}$, independent of the DM mass, as shown in Fig.~\ref{fig:kinematics}. The size of the first Brillouin zone (BZ) in Si is $2\pi/a \simeq \SI{2}{\keV}$, where $a$ is the lattice constant. Thus, for $\omega\gtrsim\SI{2}{\eV}$, DM is probing interatomic distances rather than delocalized electrons, and the electrons may be modeled as a FEG with an effective $k_F \simeq 2\pi/a$ set by the total valence electron density. This approximation is an excellent match to both density functional theory (DFT) calculations~\cite{Knapen:2020aky} and data \cite{weissker2010dynamic} for $q \simeq \SI{5}{\keV}$ and $\omega \gg E_g$ in Si \cite{walter1972frequency}; for sufficiently large $q$ ($\sim\SI{15}{\keV}$, see SM), the bound electron orbitals give large-momentum tails not captured by the FEG. 

\Cref{eq:THEANSWER} and Fig.~\ref{fig:kinematics} show that at fixed $\omega$, the rate receives contributions from $\mathcal{W}(\bb q, \omega)$ over many orders of magnitude in $q$ for $m_\chi \gtrsim \SI{10}{\mega\eV}$, so the FEG approximation is best for a light mediator, where $V(q) \propto q^{-4}$ weights the integrand most toward small $q$. Our formalism thus suggests a generic explanation for the behavior of the DM--electron spectrum in the $5\mbox{--}\SI{15}{\eV}$ range (2--4 electron--hole pairs in Si~\cite{Essig:2015cda}) from light mediator exchange in any conventional semiconductor. The projected reach in Si under the FEG approximation with a $2e^-$ threshold is shown in \cref{fig:reach} for a light  mediator.

The differences among various targets become most apparent when $\omega \simeq E_g$, where the band structure describing delocalized electrons with $q \lesssim 2\pi/a$ becomes important. In addition to band structure effects, there is also an irreducible contribution from the plasmon \cite{kundmann1988study}, where the tail extends into the kinematically allowed region for DM. This has important implications for rate predictions in currently-operating semiconductor detectors~\cite{Amaral:2020ryn,Barak:2020fql,Arnaud:2020svb}. DFT calculations predict a rate which peaks in the 1- or 2-electron bin, corresponding to $\omega \lesssim \SI{8.3}{\eV}$, for all DM masses for which these energies are kinematically accessible~\cite{Essig:2015cda}. Currently available measurements of $\mathcal{W}$ suggest the true rate in these few-electron bins may be somewhat larger. Near-gap effects are quite difficult to model~\cite{walter1972frequency}, but in our formalism, they can be accounted for by making more precise measurements at $\omega \simeq E_g$ and $q \simeq E_g/v_\chi$.

On the other hand, for near-gap scattering in a narrow-gap semiconductor ($E_g \sim \SI{10}{\meV}$), we have $q_{\mathrm{min}} \simeq \SI{10}{\eV} \ll 2\pi/a$, so the delocalized electrons in the uppermost valence band dominate the behavior of the scattering rate as $q \to 0$. We may understand the absence of screening in these systems through the Lindhard form of the dielectric function \cite{dressel2002electrodynamics}, which shows that $\epsilon(\bb q, \omega)$ has a finite limit as $\bb q \to 0$, with the imaginary part proportional to the interband transition matrix element. The lack of mobile charge carriers inhibits the screening present in metals. In the next section, we discuss an example of such a narrow-gap semiconductor: a Dirac material.

\section{Novel Materials}

Our formalism suggests that optimal materials for sub-GeV DM detection will have a loss function with large support for $\omega < v_{\chi}q$ (\cref{fig:kinematics}). For an ordinary metal with an electron effective mass $m^* = m_e$, the loss function is maximized at large $q$ when $\omega = q v_F$, where $v_F= k_F/m^{*} \simeq 10 v_\chi$. This is outside of the kinematically-allowed region for DM scattering. For small $q$, collective modes such as the plasmon will dominate, but the plasmon is damped at momenta $q > q_c \simeq \omega_p/v_F$ \cite{dressel2002electrodynamics} due to decay into the particle-hole continuum. Therefore, DM can only excite the undamped plasmon if $v_\chi > v_F$~\cite{Kurinsky:2020dpb}. Here we explore two qualitatively different ways to achieve  $v_F < v_\chi$: Dirac materials, in which $v_F$ is not tied directly to free-electron properties, and heavy-fermion materials, where strongly-correlated electrons can create a Fermi surface with a large $m^*$.
    
Dirac materials, characterized by linear electronic dispersion $\omega(k) = v_F k$ with widely-varying $v_F$ across materials \cite{doi:10.1080/00018732.2014.927109}, are promising targets for DM detection \cite{Hochberg:2017wce,Geilhufe:2018gry,Coskuner:2019odd,Geilhufe:2019ndy}. Consider a gapless isotropic Dirac material with a single Dirac cone and effective background dielectric constant $\kappa \equiv \Re[\epsilon(0,0)]$. In typical materials, $\Re(\epsilon) \gg \Im(\epsilon)$ over the relevant $\bb q$ and $\omega$ \cite{Coskuner:2019odd}, and we may write the loss function as
\begin{equation}
    \label{eq:SDirac}
    \mathcal{W}_{\mathrm{Dirac}}(q, \omega) =
        \frac{e^2}{12 \pi \kappa^2 v_F}
        \Theta(\omega - v_F q) \Theta(\omega_{\mathrm{max}} - \omega)
    .
\end{equation}
The loss function with a gap $2\Delta$ is given in the SM; $\mathcal{W}_{\rm Dirac}(q, \omega)$ is constant as $q \to 0$ for all $\omega > 2\Delta$, as anticipated. The loss function immediately displays two key features of scattering in Dirac materials \cite{Hochberg:2017wce}: small $v_F$ increases the rate, and scattering is forbidden if $v_\chi < v_F$ for $\omega = \omega_{\bb q}$. In~\cref{fig:reach}, we show the sensitivity of an isotropic Dirac material for a light mediator.

This analysis neglects many-body effects, including the plasmon contribution to the loss function. Dirac materials are expected to exhibit two tuneable plasmon modes distinct from the ordinary valence plasmon: a temperature-dependent mode which could lie in the $\mathcal{O}(\SI{}{\meV})$ range \cite{kharzeev2015universality,hofmann2015plasmon,Jenkins_2016}, and a zero-temperature mode tuneable with chemical potential \cite{thakur2017dynamical}. Therefore, measurements of the loss function in real materials are crucial to accurately estimate the scattering rate, since the plasmon contribution may dominate~\cite{kozii2018thermal} as was the case for superconductors.

Another way to lower $v_{F}$ is to find materials with ordinary quadratic dispersion but large effective masses. As an example, a number of materials containing $f$-electrons are known as heavy-fermion systems because they display a Fermi surface with $m^* \sim (10\mbox{--}100)m_e$ \cite{RevModPhys.56.755,riseborough2000heavy,coleman2015heavy}. These materials are expected to have a plasmon at energy $\omega^*_p \simeq T^*$, the Fermi temperature of the heavy electrons~\cite{millis1987plasma}. One such material is \URuSi, a heavy-fermion superconductor with $T^* = \SI{75}{\kelvin} = \SI{6.5}{\meV}$ and $m^* \simeq 6m_e$~\cite{bareille2014momentum}, from which one may estimate $v_F \simeq 6.5 \times 10^{-5}$, $\omega^*_p \simeq T^* = \SI{6.5}{\meV}$, and $q_c \simeq \omega_p^*/v_F \simeq \SI{100}{\eV}$. In reality, the measured loss function in \URuSi\ \cite{bachar2016detailed} shows considerable anisotropy with Lorentzian peaks at either \SI{4}{\meV} or \SI{6}{\meV} depending on the direction of $\bb q$, as well as a broad peak around \SI{18}{\meV}, which can also be interpreted as a heavy-fermion plasmon (see SM). Despite the extremely rich electron dynamics in this material, in our formalism we may compute the DM rate unambiguously once $\mathcal{W}$ is measured in the relevant kinematic regime.

The measured data (see SM,~\cref{fig:measuredloss}) show that $\mathcal{W}(\omega) \propto \omega$ above the heavy-fermion plasmon peaks, consistent with the tail of the ordinary valence electron plasmon. However, in contrast to spectra from conventional superconductors or semiconductors, the measured loss function in \URuSi\ shows rich structure which could be used to separate signals from backgrounds not due to fast-particle scattering. Integrating over $\omega$ from a threshold of \SI{1}{\meV} up to $\omega_{\mathrm{max}} = \SI{74}{\meV}$, the maximum value where data exists, we obtain the projected reach in \cref{fig:reach}. The band spans measurements of $\mathcal{W}(\bb q, \omega)$ as $\bb q \to 0$ along two different crystal axes. We leave a full analysis of the anisotropic response to future work~\cite{future}. As expected, the reach in \URuSi\ can surpass Al in the mass range 5--\SI{40}{\keV}, where the DM kinetic energy is comparable to the heavy-fermion plasmon energies. Our reach estimates motivate further study of \URuSi\ and similar materials as targets for light DM scattering.

\section{Implications for experiments}
The advantage of our formulation of the DM scattering rate is that no theoretical input from {\it e.g.}\ DFT is required to compute the scattering rate; the DM energy loss spectrum from spin-independent electron scattering may be precisely predicted from a measurement with an electromagnetic probe in the appropriate kinematic regime. For MeV--GeV DM, X-ray scattering covers the regime $q \sim\SI{}{\keV}$ and $\omega \sim\SI{}{\eV}$ ~\cite{schulke2007electron}, while for keV--MeV DM, momentum-resolved electron energy loss spectroscopy (EELS) can cover $q \sim\SI{}{\eV}$ and $\omega \sim\SI{}{\meV}$~\cite{raether2006excitation,osti_1530104}. These techniques are standard in condensed matter physics, and a rich literature on measurements of dielectric and loss functions already exists for a number of systems of interest. 

The downside of this formalism is that it does not directly predict how many electron--hole pairs are created in the material per unit deposited energy, or how the energy is down-converted from plasmon excitations to charge and phonons. However, if individual quasiparticle contributions to $\mathcal{W}(\bb q, \omega)$ can be modeled, this information can be reconstructed. (For related work in the context of superconducting targets, see Ref.~\cite{Hochberg:2021ymx}.) Moreover, the quasiparticle contributions may be determined empirically by correlating scattering events using an electromagnetic probe with the partition of excitations read out by the detector, as has been done for nuclear recoil calibrations at higher energy. We argue that these measurements should be considered the primary calibration mechanisms for DM--electron scattering, analogous to photoabsorption for bosonic DM absorption~\cite{Hochberg:2016ajh,Hochberg:2016sqx,Bloch:2016sjj}.

Finally, our work may be applied to unify the electronic and phonon descriptions of DM scattering with other sub-gap loss mechanisms that have not yet been explored, such as dielectric heating in insulators or coherent scattering off the superconducting condensate. Dielectric skin depth in the long-wavelength limit $\bb q \to 0$ is proportional to $\sqrt{\Re[{\epsilon(\omega)}]}/\{\omega \Im[\epsilon(\omega)]\}$, and thus materials with a small skin depth for THz photons and calorimetric readout should respond efficiently to DM--electron scattering, even for meV-scale energy deposits below the eV-scale electronic band gaps. Many materials have THz absorption features, so high-resolution THz or infrared transmission spectra are likely fertile ground for exploring new materials for keV-scale DM scattering.

\medskip

\emph{Note added.} This work appeared simultaneously with Ref.~\cite{Knapen:2021run}, which also discusses the loss function as a tool for predicting DM scattering rates. Our main results with respect to the loss function are substantively similar, although Ref.~\cite{Knapen:2021run} emphasizes comparisons with \emph{ab initio} methods, whereas the present work emphasizes the utility of the loss function formalism in target selection for future experiments.

\medskip

\emph{Acknowledgments.}  We thank Carlos Blanco, Vinayak Dravid, Rouven Essig, Sin\'{e}ad Griffin, Adolfo Grushin, David Huse, Simon Knapen, Belina von Krosigk, Jonathan Kozaczuk, Eric David Kramer, Tongyan Lin, Mariangela Lisanti, Andrea Mitridate, Lucas Wagner, and Kathryn Zurek for enlightening discussions. YK is indebted to Peter Abbamonte for relentlessly (and correctly!) emphasizing the importance of the loss function and plasmon excitations for dark matter scattering. The idea for this work was conceived via an email exchange during the “New Directions in Light Dark Matter” workshop at Fermilab, supported by
the Gordon and Betty Moore Foundation and the American Physical Society, and via a Skype call taken during the workshop ``Quantum Information and Systems for Fundamental Physics'' at the Aspen Center for Physics,  which is supported by National Science Foundation grant PHY-1607611. This project was supported in part by the Fermi National Accelerator Laboratory, managed and operated by Fermi Research Alliance, LLC under Contract No. DE-AC02-07CH11359 with the U.S. Department of Energy, through the Office of High Energy Physics QuantISED program. The work of YH is supported by the Israel Science Foundation (grant No. 1112/17), by the Binational Science Foundation (grant No. 2016155), by the I-CORE Program of the Planning Budgeting Committee (grant No. 1937/12),  and  by the Azrieli Foundation. The work of YK is supported in part by DOE grant DE-SC0015655. Parts of this document were prepared by NK using the resources of the Fermi National Accelerator Laboratory (Fermilab), a US Department of Energy, Office of Science, HEP User Facility. Fermilab is managed by Fermi Research Alliance, LLC (FRA), acting under Contract No. DE-AC02-07CH11359. The work of BVL is supported in part by DOE grant DE-SC0010107. TCY is supported by the U.S. Department of Energy under contract number DE-AC02-76SF00515. KKB acknowledges support for the later stages of the work from the Fermi Research Alliance, LLC (FRA) and the US Department of Energy (DOE) under contract No. DE-AC02-07CH11359; the initial stages of the work were supported by the DOE under the QuantiSED program, Award No. DE-SC0019129.

\bibliography{DielectricRefs}
\onecolumngrid
\clearpage

%%%%%%%%%% Supplemental materials %%%%%%%%%%
\setcounter{page}{1}
\setcounter{equation}{0}
\setcounter{figure}{0}
\setcounter{table}{0}
\setcounter{section}{0}
\setcounter{subsection}{0}
\renewcommand{\theequation}{S.\arabic{equation}}
\renewcommand{\thefigure}{S\arabic{figure}}
\renewcommand{\thetable}{S\arabic{table}}
\renewcommand{\thesection}{\Roman{section}}
\renewcommand{\thesubsection}{\Alph{subsection}}
\newcommand{\ssection}[1]{
    \addtocounter{section}{1}
    \section{\thesection.~~~#1}
    \addtocounter{section}{-1}
    \refstepcounter{section}
}
\newcommand{\ssubsection}[1]{
    \addtocounter{subsection}{1}
    \subsection{\thesubsection.~~~#1}
    \addtocounter{subsection}{-1}
    \refstepcounter{subsection}
}
\newcommand{\fakeaffil}[2]{$^{#1}$\textit{#2}\\}

\thispagestyle{empty}
\begin{center}
    \begin{spacing}{1.2}
        \textbf{\large 
            Supplemental Material:\\Determining Dark Matter-Electron Scattering Rates from the Dielectric Function
        }
    \end{spacing}
    \par\smallskip
    Yonit  Hochberg,$^1$
    Yonatan  Kahn,$^{2,\,3}$
    Noah Kurinsky,$^{4,\,5}$
    Benjamin\\V. Lehmann,$^{6,\,7}$
    To Chin Yu,$^{8,\,9}$
    and Karl  K.  Berggren$^{10}$
    \par
    {\small
        \fakeaffil{1}{Racah Institute of Physics, Hebrew University of Jerusalem, Jerusalem 91904, Israel}
        \fakeaffil{2}{Department of Physics, University of Illinois at Urbana-Champaign, Urbana, IL 61801, USA}
        \fakeaffil{3}{Illinois Center for Advanced Studies of the Universe, University of Illinois at Urbana-Champaign, Urbana, IL 61801, USA}
        \fakeaffil{4}{Fermi National Accelerator Laboratory, Batavia, IL 60510, USA}
        \fakeaffil{5}{Kavli Institute for Cosmological Physics, University of Chicago, Chicago, Illinois 60637, USA}
        \fakeaffil{6}{Department of Physics, University of California Santa Cruz, Santa Cruz, CA 95064, USA}
        \fakeaffil{7}{Santa Cruz Institute for Particle Physics, Santa Cruz, CA 95064, USA}
        \fakeaffil{8}{Department of Physics, Stanford University, Stanford, CA 94305, USA}
        \fakeaffil{9}{SLAC National Accelerator Laboratory, 2575 Sand Hill Road, Menlo Park, CA 94025, USA}
        \fakeaffil{10}{Massachusetts Institute of Technology, Department of Electrical Engineering and Computer Science, Cambridge, MA 02139, USA}
    }
\end{center}
\par\smallskip

In this Supplemental Material, we provide a number of derivations and further details to support the results in the main Letter. \Cref{sec:s-scattering-rate} derives our main result for the DM scattering rate in terms of the loss function. \Cref{sec:s-loss-function} outlines a number of simple analytic models for dielectric functions in various materials, and compares them to measured data for Al (representative of an ordinary superconductor), Si (a typical semiconductor), and \URuSi\ (an example of a heavy-fermion superconductor with meV-scale plasmons). We compare the free-electron gas (FEG) model for Si to the spectrum computed using crystal form factors generated by the publicly-available \texttt{QEdark} code \cite{Essig:2015cda}, and show good qualitative agreement in the range 5--\SI{15}{\eV}. \Cref{sec:s-reach-projections} is devoted to a detailed comparison of our results for superconductors with other results in the literature, justifying our claim of a stronger reach by several orders of magnitude compared to previous estimates, and gives the projected reach for heavy mediators.

\ssection{Scattering rate in terms of the loss function}
\label{sec:s-scattering-rate}

Here we derive \cref{eq:THEANSWER} and show how the scattering rate for all spin-independent DM-electron interactions is governed by the loss function. Suppose DM couples to electrons through a low-energy Hamiltonian of the form
\begin{equation}
    \label{eq:HDMSM}
    \hat{H}_{\mathrm{int}} = \sum_{i} V(\rhat_\chi - \rhat_i)
    ,
\end{equation}
where the sum runs over all electrons in the target. Fourier transforming the potential,
\begin{equation}
    V(\rhat_\chi - \rhat_i) = \int \frac{\du^3\bb q}{(2\pi)^3}
        e^{i \bb q \cdot(\rhat_\chi - \rhat_i)}V(\bb q)
    ,
\end{equation}
we can write the interaction Hamiltonian as
\begin{equation}
    \hat{H}_{\mathrm{int}}= 
    \int \frac{\du^3\bb q}{(2\pi)^3}  e^{i \bb q \cdot \rhat_\chi}
    V(\bb q) \hat{\rho}(\bb q)
    ,
\end{equation}
where the momentum-space electron density operator is defined as
\begin{equation}
    \hat{\rho}(\bb q) = \int\du^3\bb x \sum_{i} \delta(\bb x - \rhat_i)
        e^{-i \bb q \cdot \bb x} = \sum_i e^{-i \bb q \cdot \rhat_i}
    .
\end{equation}
By Fermi's Golden Rule (equivalently, the Born approximation), we can compute the transition rate $\Gamma(\bb v_\chi)$ from the ground state $|0 \rangle$ for a given incoming DM velocity $\bb v_\chi$, treating the incoming and outgoing DM as plane waves with energy and momentum $(E_\chi, \mathbf{p}_\chi)$ and $(E'_\chi, \mathbf{p}_\chi')$ respectively. We take the ground state to have zero energy without loss of generality. The transition rate is given by \cite{Trickle:2019nya}
\begin{align}
\Gamma(\bb v_\chi)
    & = \sum_f \bigl|\langle
        f;  \mathbf{p}_\chi'  | \hat{H}_{\mathrm{int}} | 0; \mathbf{p}_\chi
    \rangle\bigr|^2 2\pi \delta(\omega_f + E_\chi' - E_\chi )
    \nonumber\\
    \label{eq:FGR}
    & = \int \frac{\du^3\bb q}{(2\pi)^3} |V(\bb q)|^2
        \sum_f \bigl| \langle f| \hat{\rho}(\bb q) | 0 \rangle \bigr|^2
        2\pi \delta (\omega_f - \omega_{\bb q})
    ,
\end{align}
where $|f \rangle$ is a final state with energy $\omega_f$ and the sum runs over all possible final states of the system, and we recall that
\begin{equation}
    \omega_{\bb q} = \bb q \cdot \bb v_\chi - \frac{q^2}{2m_\chi}.
\end{equation}
Note that the only assumption that was made here was that $\hat{H}_{\mathrm{int}}$ is sufficiently weak compared to the unperturbed Hamiltonian $\hat{H}_0$ of the target system; this is the case in Ref.~\cite{nozieres1959electron} for ordinary electron-electron scattering, so it must be the case for DM-electron scattering where the couplings are much weaker. Note this implies one cannot directly apply our result to regions of parameter space where DM and electrons are strongly coupled, as would be relevant for regions in parameter space where DM may not reach underground detectors due to multiple scattering \cite{Kavanagh:2016pyr,Emken:2017erx,Emken:2017qmp,Emken:2018run,Emken:2019tni}. 

The insight of Ref.~\cite{nozieres1959electron} is to relate the density matrix element $| \langle f | \hat{\rho}(\bb q) | 0 \rangle |^2 $ to an experimentally measurable quantity, the dielectric function $\epsilon(\bb q, \omega)$. The dielectric function is \emph{defined} as the linear response of the target to the longitudinal electric field of a test charge. For simplicity and to elucidate the formalism, in this work we consider the case of an isotropic material where the dielectric function is a scalar rather than a tensor, and relegate the treatment of the anisotropic case to upcoming work~\cite{future}. Since a test charge will also perturb the electron density of the target, it can be shown that this is equivalent to defining the dielectric function as a density-density correlation function \cite{altland2010condensed}. Of course, the electrons will also couple to ions, and strictly speaking the ion density operator should also appear in the measured loss function. In what follows, we will assume that these contributions are negligible, which is the approximation always made in the condensed matter literature. This assumption makes our formalism independent of the DM coupling to protons or neutrons, and even if ion contributions are significant, the measured loss function will give \emph{exactly} the correct rate for a dark photon mediator.

Because the dielectric function is defined as the \emph{linear} response of the system, the same assumptions are implicit in the setup of \cite{nozieres1959electron} (with an electromagnetic probe) as are present in the DM scattering setup: the test charge interactions are weak compared to the internal interactions $\hat{H}_0$. Therefore the Coulomb potential of the test charge may be factored out in Fourier space, separating the (weak) perturbation due to the probe and the (possibly strong) response of the system to such a probe.  The result is \cite{nozieres1959electron, altland2010condensed}
\begin{equation}
    \label{eq:lossfunction}
    \Im\left(-\frac{1}{\epsilon(\bb q, \omega)}\right) =
        \frac{\pi e^2}{q^2} \sum_f
        \bigl| \langle f | \hat{\rho}(\bb q) | 0 \rangle \bigr|^2
        \delta(\omega_f - \omega)
    .
\end{equation}
Here we are using Heaviside-Lorenz conventions for the electron charge $e$ as is common in high-energy physics, which differs from the Gaussian unit definition common in condensed matter physics by a factor of $\sqrt{4\pi}$. (Note also that Eq.~(9) of \cite{nozieres1959electron} is missing a factor of $\pi$.) Plugging \cref{eq:lossfunction} into \cref{eq:FGR}, we obtain our main result, \cref{eq:THEANSWER}. For reference, the sum over final states on the right-hand side of \cref{eq:lossfunction} is (up to factors of $\pi$) conventionally defined in the condensed matter literature as the dynamic structure factor. 

Notice that we have made no assumptions whatsoever about the character of the final state $|f \rangle$. It is an exact eigenstate of the (in general very complicated) many-body condensed matter Hamiltonian, and the only requirement is that it represents some rearrangement of the electrons in the target so that it has a nonzero matrix element with the electron density operator with respect to the ground state. In this sense our treatment is distinct from Ref.~\cite{Trickle:2019nya}, which defines a general dynamic structure factor (with slightly different normalization compared to the condensed matter conventions) very similar to the sum in \cref{eq:FGR}, but one which is excitation-specific and requires quantization in terms of single-quasiparticle states in the case of electron scattering. When all many-body states are included, the structure factor defined in Ref.~\cite{Trickle:2019nya} for electron scattering is \emph{identical} to the loss function defined through the complex dielectric function, is directly measurable without the need to compute single-particle wavefunctions, and automatically includes all in-medium effects. (Our formalism is philosophically similar to Ref.~\cite{Banks:2020gpu}, which parameterizes non-relativistic potentials using only general principles such as the K\"{a}ll\'{e}n-Lehmann spectral representation, without relying on the assumption of the perturbative exchange of a single mediator.) On the other hand, the formalism of Ref.~\cite{Trickle:2019nya} is useful when DM couples differently to electrons, protons, and neutrons than the photon, in an energy regime where density perturbations to both electrons and ions are relevant, as might be the case for sub-gap single-phonon excitations. 

Finally, we note that other UV Lagrangians considered in Ref.~\cite{Trickle:2020oki} also generate non-relativistic potentials which couple to the electron density, but are often accompanied by other spin- or momentum-dependent operators which may complicate our arguments, so we focus on the case of spin-independent scattering. In particular, if DM is a Dirac fermion $\chi$ which couples to a scalar $\phi$ of mass $m_\phi$ through the scalar current $\mathcal{L} \supset g_\chi \phi \bar{\chi}\chi$, or to a vector $V_\mu$ of mass $m_V$ through the vector current $\mathcal{L} \supset g_\chi V_\mu \bar{\chi} \gamma^\mu \chi$, and if the mediator couples to electrons in an analogous fashion but with coupling $g_e$, the resulting potential is the same in both cases \cite{Trickle:2020oki}:
\begin{equation}
    \label{eq:potentialSM}
    V(\bb q) = V(q) = \frac{g_\chi g_e}{q^2 + m^2_{\phi,V}}.
\end{equation}
Similar formulas apply when DM is a complex scalar. Note that in contrast with Ref.~\cite{Trickle:2020oki}, we leave the DM-electron coupling as its `bare' value and place all in-medium corrections to this coupling entirely within the loss function. In the case where $g_e \propto e$, as would be the case for a kinetically-mixed dark photon mediator or when the DM is millicharged, the factors of $1/e^2$ cancel in \cref{eq:THEANSWER} because the DM-induced perturbation to the electron density is exactly proportional to an ordinary electromagnetic probe. 

For completeness, we give the expression for the energy spectrum from DM-electron scattering,
\begin{equation}
    \label{eq:spectrum}
    \frac{\du R}{\du\omega} = \frac{\rho_\chi}{2 \pi^2 e^2 \rho_T m_\chi}
        \int\du q \, q^3 |V(q)|^2 \mathcal{W}(q,\omega)
        \eta\bigl(v_{\mathrm{min}}(q,\omega)\bigr)
    ,
\end{equation}
where $\rho_T$ is the mass density of the target, $\eta(v_{\mathrm{min}})$ is the mean inverse DM speed $\int_{v_{\mathrm{min}}} \du ^3 \bb v_\chi f(\bb v_\chi)/v_\chi$, and $v_{\mathrm{min}} = \frac{\omega}{q} + \frac{q}{2m_\chi}$ is the minimum DM speed required to produce an excitation with momentum $q$ and energy $\omega$ for DM of mass $m_\chi$. To compare with the literature, we take $f(\bb v_\chi)$ to be the standard halo model with dispersion $v_0 = \SI{220}{\kms}$, escape velocity $v_{\mathrm{esc}} = \SI{550}{\kms}$, and Earth velocity $v_E = \SI{232}{\kms}$ in the galactic frame. Integrating \cref{eq:spectrum} over $\omega$ within the dynamic range of a given experiment gives the total scattering rate.

\ssection{Models and measurements of the loss function}
\label{sec:s-loss-function}

In our formalism, the detector response and its influence on the scattering rate are entirely captured by the complex dielectric function $\epsilon(\bb q, \omega)$ via the loss function $\mathcal W$ of \cref{eq:lossfunction} and \cref{eq:Wdef}. In principle, this quantity is directly measurable with electromagnetic probes in a given material. However, most measurements presently available in the literature are made at values of $(\bb q, \omega)$ different than those of interest for the detection of light DM (see \cref{fig:kinematics}). Thus, for a first estimate of the scattering rate, we employ analytical approximations to the dielectric function. Important consistency checks can be implemented based on the fact that $\epsilon^{-1}$ is defined as a causal correlation function, and thus must have certain analytic properties. In particular, the following two `sum rule' relations are satisfied exactly by $\mathcal{W}(\bb q, \omega)$ in the limit of an isotropic system \cite{mahan2013many}:
\begin{align}
    \label{eq:sum-rule-1}
    \int_0^\infty\du\omega \, \omega \, \mathcal{W}(\bb q, \omega) =
        \frac{\pi}{2}\omega_p^2
    , \\
    \label{eq:sum-rule-2}
    \lim_{\bb q\to 0}\int_0^\infty\du\omega\,
        \frac{\mathcal{W}(\bb q, \omega)}{\omega} = \frac{\pi}{2}.
\end{align}
\Cref{eq:sum-rule-1} is effectively a manifestation of charge conservation, which explains the appearance of the plasma frequency
\begin{equation}
\omega_p^2 = \frac{4 \pi \alpha n_e}{m_e},
\end{equation}
which is proportional to the total electron density $n_e$ in the FEG limit, while \cref{eq:sum-rule-2} follows from causality. Causality also implies that $\mathcal{W}(\bb q, -\omega) = - \mathcal{W}(\bb q, \omega)$ \cite{mahan2013many}, which has important consequences for the projected reach in superconductors, as we will see below.

\ssubsection{RPA dielectric function for a homogeneous electron gas}

An analytic form for the dielectric function of a homogeneous electron gas can be derived from first principles under the random phase approximation~(RPA). Here a word about terminology is in order: screening effects arise from Coulomb interactions between electrons, but in RPA these are embodied in the total scalar potential for the system which is solved for self-consistently~\cite{mahan2013many}. Thus RPA captures only a certain subset of electron interactions without including electron-electron interactions directly in the Hamiltonian; in QFT language, it sums the series of ladder diagrams constructed from the 1-loop vacuum polarization to obtain the resummed photon propagator, but does not include higher-loop diagrams involving additional electron lines. This is the sense in which the electrons are treated as `free' and $\epsilon_{\mathrm{RPA}}$ is sometimes referred to as the dielectric function for the free electron gas (FEG). Below we will consider further improvements to this approximation. 

The resulting dielectric function at zero temperature is given by Eq.~(5.4.21) of Ref.~\cite{dressel2002electrodynamics} as
\begin{multline}
    \label{eq:e-rpa-fw}
    \epsilon_{\mathrm{RPA}}(\bb q,\omega) =
    1 + \frac{3\omega_p^2}{q^2v_F^2}\Biggl\{
        \frac12 + \frac{k_F}{4q}\left(
            1 - \left(
                \frac{q}{2k_F} - \frac{\omega + i\Gamma_p}{qv_F}
            \right)^2
        \right)\operatorname{Log}\left(
            \frac{\frac{q}{2k_F}-\frac{\omega + i\Gamma_p}{qv_F} + 1}
                 {\frac{q}{2k_F}-\frac{\omega + i\Gamma_p}{qv_F} - 1}
        \right) \\
        +
        \frac{k_F}{4q}\left(
            1 - \left(
                \frac{q}{2k_F} + \frac{\omega + i\Gamma_p}{qv_F}
            \right)^2
        \right)\operatorname{Log}\left(
            \frac{\frac{q}{2k_F}+\frac{\omega + i\Gamma_p}{qv_F} + 1}
                 {\frac{q}{2k_F}+\frac{\omega + i\Gamma_p}{qv_F} - 1}
        \right)
    \Biggr\}
    .
\end{multline}
Here $\operatorname{Log}$ denotes the principal value of the natural logarithm, $k_F$ and $v_F$ are the Fermi momentum and Fermi velocity respectively, and $\Gamma_p$ is a free parameter controlling the width of the plasmon which can also be interpreted as a quasiparticle lifetime. The plasma frequency can also be written in the form
\begin{equation}
    \omega_p = \frac{\lambda_{\mathrm{TF}}v_F}{\sqrt{3}}
        = \frac{v_F}{\sqrt3}
            \left[\frac{e}{\pi}\left(2E_Fm_e^3\right)^{1/4}\right]
    ,
\end{equation}
where $\lambda_{\mathrm{TF}}$ is the (inverse) Thomas--Fermi screening length. We expect the zero-temperature RPA result to be an excellent approximation for $\omega \gg 2\Delta$, where $2\Delta$ is the superconducting gap. As mentioned in the main text, this approximation ignores possible enhancements to the loss function from scattering off of the condensate at energies near or below the gap, which will be considered in future work~\cite{future}. In the literature, \cref{eq:e-rpa-fw} is known as the Lindhard dielectric function, though Lindhard's formalism may also be applied to semiconductors as well as metals; in what follows, we will use the terms `Lindhard,' `RPA,' and `FEG' interchangeably to refer to \cref{eq:e-rpa-fw}.

Observe that the arguments of the logarithms in \cref{eq:e-rpa-fw} are in general complex. For some values of $q$ and $\omega$, these arguments lie along the negative real axis in the narrow-width limit $\Gamma_p\to0$, and the imaginary part of $\epsilon$ then depends crucially on the choice of branch. The branch choice is fixed by the causality condition $\mathcal{W}(\bb q, -\omega) = - \mathcal{W}(\bb q, \omega)$, which is automatic for positive real values of $\Gamma_p$, but the $\Gamma_p\to0$ limit is non-trivial. The causal result is given by Eq.~(5.4.22b) of Ref.~\cite{dressel2002electrodynamics} as
\begin{align}
    \label{eq:re-r-rpa}
    &\Re\epsilon_{\mathrm{RPA}}(\bb q,\omega) \simeq
        1 + \frac{\lambda_{\mathrm{TF}}^2}{q^2}\left(
            \frac12 + \frac{k_F}{4q}\left(
                1 - Q_-^2
            \right)\log\left|\frac{Q_- + 1}{Q_- - 1}\right| +
            \frac{k_F}{4q}\left(
                1 - Q_+^2
            \right)\log\left|\frac{Q_+ + 1}{Q_+ - 1}\right|
        \right)
    ,
    \\
    \label{eq:im-e-rpa}
    &\Im\epsilon_{\mathrm{RPA}}(\bb q,\omega) \simeq
    \frac{3\pi\omega_p^2}{q^3v_F^2}\begin{cases}
        \omega/(2 v_F)
            & Q_+ < 1 \\
        k_F\left(1 - Q_-^2\right)/4
            & \left|Q_-\right| < 1 < Q_+ \\
        0 & \left|Q_-\right| > 1.
    \end{cases}
\end{align}
where $Q_\pm = \frac{q}{2k_F} \pm \frac{\omega}{qv_F}$. The acausal branch prescription was employed in Ref.~\cite{Hochberg:2015fth}, which as we will see in \cref{sec:s-reach-projections} below, artificially suppresses the scattering rate for low DM masses.

The imaginary part of the Lindhard dielectric function naturally contains the plasmon as a Lorentzian peak at $\omega=\omega_p$ of width $\Gamma_p$. For the purposes of light DM detection, kinematics favor energy deposits $\omega\ll\omega_p$. The plasmon has then typically been neglected in the literature in the computation of the scattering rate, {\it i.e.}, the rate is computed in the limit $\Gamma_p\to0$. However, for realistic values of $\Gamma_p$, the tail of the plasmon peak may significantly contribute to or even dominate the loss function at the relevant values of $\omega$.

For DM--electron scattering in semiconductors, if the deposited energy is $\mathcal{O}(\SI{5}{\eV})$ or greater, the minimum momentum transfer is $q \gtrsim \SI{5}{\keV}$ independent of the DM mass (see \cref{fig:kinematics} in the main text). Since $k_F \simeq 2\pi/a \simeq \SI{5}{\keV}$ for typical interatomic spacings $a$, this means that the behavior of this part of the spectrum will be determined by the loss function in the region $q  \gtrsim k_F$. For these values of $q$, the DM is probing length scales smaller than the distance between lattice sites, so we might expect that the inhomogeneities due to the lattice become unimportant and the response is similar to a FEG. For $q > 2k_F$ the loss function peaks when $Q_- \approx 0$, corresponding to $\omega = \frac{q^2 v_F}{2k_F} = \frac{q^2}{2m_e}$, which is elastic scattering from free electrons at rest. For a given $q$, the loss function is nonzero over a range $\Delta \omega \simeq 2q v_F$ around the peak, reflecting the fact that electrons at the Fermi surface have a nonzero velocity. Note however that the loss function vanishes when $|Q_-| > 1$, which can happen for sufficiently small $\omega$ at sufficiently large $q$. This is an artificial feature of the FEG which is not present in semiconductors, where the valence (and core) electron wavefunctions have a tight-binding character with a momentum-space tail that extends to arbitrarily large values. This regime corresponds to $q \gtrsim Z_{\mathrm{eff}}/a_0 \simeq \SI{15}{\keV}$ where $a_0$ is the Bohr radius and $Z_{\mathrm{eff}} \approx 4$ is the effective nuclear charge felt by the valence electrons in Group 14 elements (carbon, silicon, and germanium). The large-$q$ behavior is especially apparent in some materials like germanium, where the $3d$ shell may become energetically accessible for $\omega$ exceeding the binding energy. A corresponding feature is seen in the spectrum in models using tight-binding wavefunctions \cite{Lee:2015qva} as well as those using density functional theory (DFT) techniques \cite{Essig:2015cda}.

\ssubsection{Plasmon pole approximation and local field corrections}

In the limit that the plasmon dominates, the dielectric function may be derived by modeling the atomic response as a damped harmonic oscillator. This is known as the Fr\"ohlich model \cite{frohlich1959phenomenological}, and the result is
\begin{equation}
    \epsilon_{\mathrm{F}}(\bb q,\omega) =
        \epsilon_{c} +
        \frac{\omega_p^2}{(\omega_g^2 - \omega^2) - i\omega\Gamma_p}
    .
\end{equation}
Here $\epsilon_{c}$ denotes the contribution from core electrons, which is assumed to be independent of $\bb q$ and $\omega$, and $\omega_g$ is an average band gap which can be set to zero for metals. The corresponding loss function features a Breit--Wigner-like peak, with the form
\begin{equation}
    \label{eq:frolich}
    \mathcal W_{\mathrm F}(\bb q, \omega) = 
        \frac{\omega_p^2\omega\Gamma_p}
             {\epsilon_c^2\left(
                \omega_g^2 + \omega_p^2/\epsilon_c^2 - \omega^2
            \right)^2 + \omega^2\Gamma_p^2}
    .
\end{equation}
This function satisfies the sum rules of \cref{eq:sum-rule-1,eq:sum-rule-2} with $\epsilon_c=1$ and $\omega_g=0$. Note that this form of the loss function is linear in $\omega$ for $\omega\ll\omega_p$.

The low-energy loss function is also subject to effects which are not included in the Lindhard dielectric function. Ref.~\cite{gibbons1976line} (hereafter denoted `GSRF') fits the plasmon in aluminum including a local-field correction and accounting for the polarizability of atomic cores $\chi_{\mathrm{core}}$, resulting in a dielectric function of the form
\begin{equation}
\label{eq:eps-gsrf}
    \epsilon_{\mathrm G}(\bb q, \omega) =
    1 + \frac{
        \left[\omega + i\Gamma_p(\bb q)\right]\left[
            \epsilon_{\mathrm{RPA}}(\bb q, \omega)
            - 1 + 4\pi\chi_{\mathrm{core}}
        \right]
    }{
        \omega\left(
            1 - G(\bb q)\left[
                \epsilon_{\mathrm{RPA}}(\bb q, \omega) - 1
            \right]
        \right) + i\Gamma_p(\bb q)\left(
            1 - G(\bb q)\left[
                \epsilon_{\mathrm{RPA}}(\bb q, 0) - 1
            \right]
        \right)\frac{
            \epsilon_{\mathrm{RPA}}(\bb q, \omega) - 1 + 4\pi\chi_{\mathrm{core}}
        }{
            \epsilon_{\mathrm{RPA}}(\bb q, 0) - 1 + 4\pi\chi_{\mathrm{core}}
        }
    }
    ,
\end{equation}
where $G(\bb q)$ is known as the exchange parameter and arises in the microscopic theory from 1-loop corrections to the electron-photon vertex \cite{mahan2013many}. Ref.~\cite{gibbons1976line} provides fits to $G$ and $\Gamma_p$ as functions of $\bb q$. Complex values of $G$ produce damping, which influences the form of the loss function at small values of $\omega$. However, $\Im G(\bb q)\neq 0$ can lead to unphysical negative values of the loss function at the smallest values of $\omega$, thereby violating the positivity requirements imposed by the sum rules, and moreover $G$ as computed in various microscopic theories tends to be real \cite{mahan2013many}. Following Ref.~\cite{gibbons1976line}, we divide our treatment into two cases, one with complex-valued $G$ (`damped') and one with real-valued $G$ (`undamped').

\ssubsection{Dielectric function for Dirac materials}

Dirac materials are characterized by electrons with the approximately linear dispersion characteristic of relativistic Dirac fermions, rather than the usual quadratic dispersion expected at a band minimum. In real materials, there are typically two such bands, one below and one above the Fermi energy, with dispersions $E_{\pm}(\bb k) = \pm \sqrt{v_F^2 \bb k^2 + \Delta^2}$. Here, $\Delta$ plays the role of the fermion mass and the Fermi velocity $v_F$ is the analogue of the speed of light; the gap at the Dirac point with $\bb k = 0$ is $2\Delta$. The band structure may be anisotropic, with different Fermi velocities along different lattice directions, but for pedagogical purposes we will focus here on isotropic materials; see Refs.~\cite{Geilhufe:2019ndy,Coskuner:2019odd} for a detailed investigation of anisotropic Dirac materials for DM detection.

In the approximation that only two nondegenerate bands contribute to the Dirac electron spectrum, the dielectric function may be computed using Lindhard's formalism in the Bloch wave basis \cite{dressel2002electrodynamics}. At zero temperature, with the valence ($-$) band full and the conduction ($+$) band empty, this reads
\begin{equation}
    \label{eq:LindhardBand}
    \epsilon_{\mathrm{Dirac}}(\bb q, \omega) = 1 + \lim_{\eta \to 0} 
    \frac{1}{V}\frac{e^2}{q^2} \int_{\mathrm{BZ}}
    \frac{V_{\mathrm{uc}}\du^3\bb k}{(2\pi)^3}
    \frac{2}{E_+(\bb k + \bb q) - E_-(\bb k) - \omega - i \eta}
    \left|\langle\bb k + \bb q; + |
        e^{i \bb q \cdot \bb r}
    | \bb k; - \rangle\right|^2
    ,
\end{equation}
 where $|\bb k; \pm \rangle$ represents a Bloch wavefunction with crystal momentum $\bb k$ in the band $-$ or $+$, $V$ is the crystal volume, the factor of 2 is for spin degeneracy, and the integral is taken over the first Brillouin zone (BZ) in the continuum limit using the unit cell volume to regularize the momentum sum, $\sum_{\bb k} \to \int V_{\mathrm{uc}}\,  \du^3\bb k/(2\pi)^3$. There are some complications with this procedure in the case of anisotropic materials \cite{Coskuner:2019odd}, but it yields an accurate estimate for the imaginary part in isotropic materials, which is dominated by the smallest gaps and hence the bands other than the Dirac bands may be neglected. However, as noted in Ref.~\cite{Coskuner:2019odd}, $\Re[\epsilon(0,0)]$ acts as a background dielectric constant receiving contributions from the entire BZ and thus cannot be reliably calculated analytically. We may therefore estimate the real part as simply $\Re(\epsilon_{\mathrm{Dirac}}) = \kappa \gg 1$ independent of $\bb q$ and $\omega$ over the relevant kinematic range.

To obtain the imaginary part, we may use the identity $\Im (\lim_{\eta \to 0}\frac{1}{x - i \eta}) = \pi \delta(x)$ and perform the integral using spinor wavefunctions with the matrix element given in Ref.~\cite{Hochberg:2017wce}. Note that this is precisely analogous to performing the phase space integral over the valence and conduction bands in the single-particle formalism for determining the scattering rate; the dielectric function allows us to express the results of Ref.~\cite{Hochberg:2017wce} in a more convenient and generalizable formalism. Equivalently, we may recognize that with the replacements $\Delta \to m_e$ and $v_F \to c$, the imaginary part is identical to that of the 1-loop vacuum polarization in relativistic quantum electrodynamics (QED), which is proportional to the cross section for $\gamma^* \to e^+ e^-$ by the optical theorem~\cite{Peskin:1995ev}. The result is
\begin{equation}
    \Im \epsilon_{\mathrm{Dirac}} (q, \omega) =
    \frac{e^2}{12 \pi v_F}\sqrt{1 - \frac{4 \Delta^2}{\omega^2 - v_F^2 q^2}}
    \left(1 + \frac{2\Delta^2}{\omega^2 - v_F^2 q^2}\right)
    \Theta(\omega^2 - v_F^2 q^2 - 4 \Delta^2),
\end{equation}
where the coefficient $e^2/(12\pi)$ is (up to a factor of $\pi$) the familiar 1-loop beta function coefficient of QED. Indeed, the physics of the dielectric function is the same in Dirac materials as it is in the true QED vacuum; the screening of bare charges due to $\Im (\epsilon)$ at $q \simeq 2m_e$ is known as the Uehling potential.

As long as $v_F$ is not too small, $\Im(\epsilon) \lesssim 1$. (Otherwise perturbation theory would break down, as noted in Ref.~\cite{Hochberg:2017wce}.) Then if $\kappa \gg 1$, we may approximate $\Im(-1/\epsilon) \approx \Im(\epsilon)/\kappa^2$ and thus
\begin{equation}
    \mathcal{W}_{\mathrm{Dirac}} (q, \omega) = \frac{e^2}{12 \kappa^2 \pi v_F}\sqrt{1 - \frac{4 \Delta^2}{\omega^2 - v_F^2 q^2}} \left(1 + \frac{2\Delta^2}{\omega^2 - v_F^2 q^2}\right) \Theta(\omega^2 - v_F^2 q^2 - 4 \Delta^2) \Theta(\omega_{\mathrm{max}} - \omega)
\end{equation}
Setting $\Delta = 0$ gives \cref{eq:SDirac} in the main text. The last factor may be explained as follows. In real materials, the Dirac band structure does not extend throughout the entire BZ, but deviates from linearity at some point. In Ref.~\cite{Hochberg:2017wce} this was expressed as a momentum cutoff $\Lambda$, which is required to regularize the real part of $\epsilon_{\mathrm{Dirac}}$. Here, since we are dealing with model functions rather than real materials, we instead impose a cutoff $\omega_{\mathrm{max}}$ on the depth of the Dirac band, which has typical values of $\omega_{\mathrm{max}} \simeq \SI{0.5}{\eV}$ in {\it e.g.} ZrTe$_5$ \cite{Hochberg:2017wce}. Finally, note that $\mathcal{W}_{\mathrm{Dirac}}$ violates the causality requirement $ \mathcal{W}_{\mathrm{Dirac}}(q, -\omega) = -\mathcal{W}_{\mathrm{Dirac}}(q, \omega)$. This indicates that $\mathcal{W}_{\mathrm{Dirac}}$ as computed here does not represent the entire loss function, and in particular (as noted in the main text) it is missing plasmon contributions.

\ssubsection{Measurements of the loss function in various materials}

\begin{figure}
    \centering
    \includegraphics[width=\columnwidth]{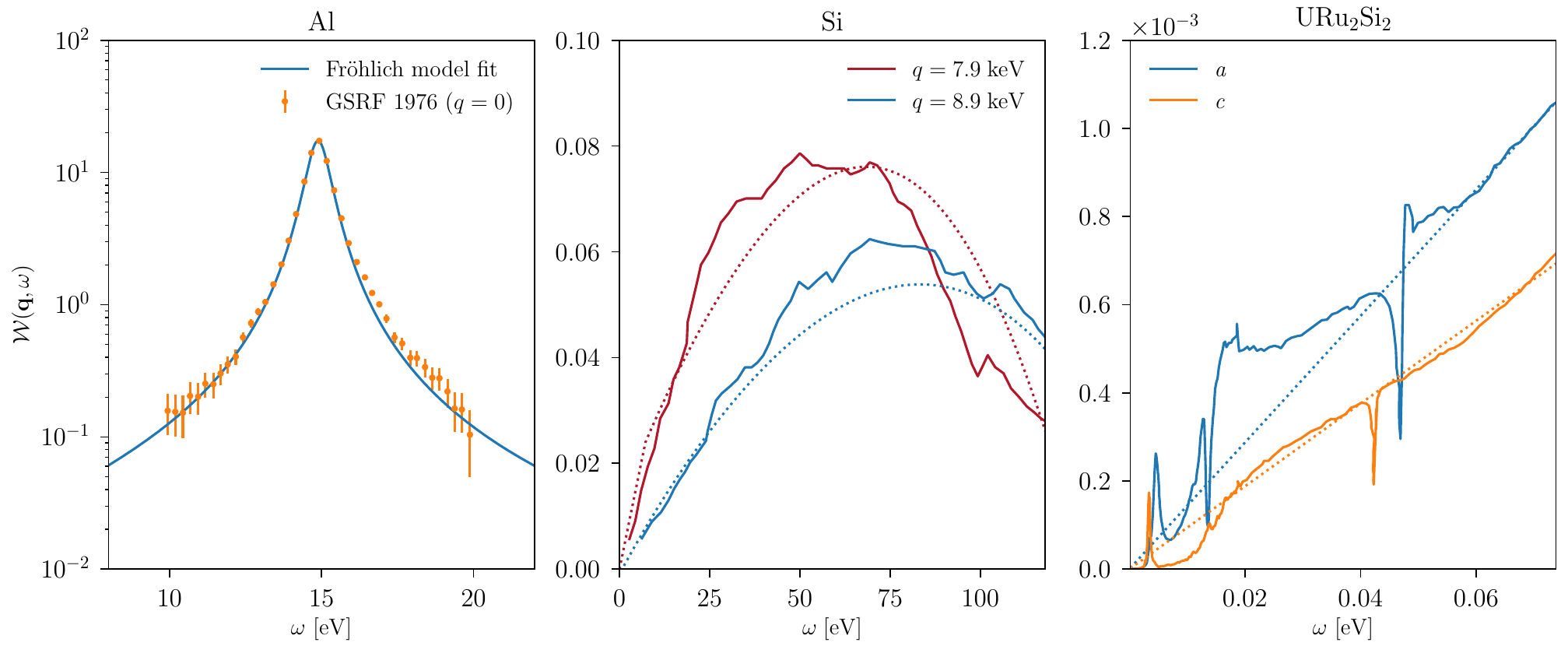}
    \caption{Measurements of the loss function $\mathcal{W}(\bb q, \omega) = \Im(-1/\epsilon(\bb q, \omega))$ in Al, Si, and \URuSi\ along with model fits when appropriate. \textbf{Left:} loss function for Al at $q = 0$ in the vicinity of the plasmon peak from Ref.~\cite{gibbons1976line}, fit with the Fr\"ohlich model of \cref{eq:frolich}. The best-fit parameters are $(\omega_p,\,\Gamma_p) = (\SI{14.9}{\eV},\;\SI{0.863}{\eV})$. Error bars indicate the accuracy with which the data points could be transcribed from Ref.~\cite{gibbons1976line}. \textbf{Center:} loss function for Si at large momenta $q > 2\pi/a$, measured from X-ray scattering in Ref.~\cite{weissker2010dynamic}. Dashed lines show the Lindhard RPA loss function with $\Gamma_p = 0$, $k_F = m_e v_F$, and $\omega_p=\SI{16.67}{\eV}$ \cite{kundmann1988study}. The Fermi velocity is treated as a free parameter and is fixed here to the best-fit value of $v_F=\SI{2.59e8}{\centi\meter/\second} = 8.6 \times 10^{-3}$ in natural units, which is comparable to Fermi velocities of metals with similar densities. \textbf{Right:} loss function in \URuSi\ at $q = 0$ measured along two different crystal axes $a$ and $c$ at $T = \SI{9}{\kelvin}$ \cite{bachar2016detailed} (solid), along with a linear fit to both datasets (dashed). If interpreted as the tail of a valence electron plasmon, the slope should be $\Gamma_p/\omega_p^2$. The fit gives a slope of  $\Gamma_p/\omega_p^2  \simeq 14\;(9) \times \SI{e-3}{\eV^{-1}}$ along the $a$ ($c$) axis which implies $\Gamma_p/\omega_p \simeq 0.21\;(0.13)$ for $\omega_p \simeq \SI{15}{\eV}$, values which are typical for other metals.}
    \label{fig:measuredloss}
\end{figure}

\begin{figure*}
    \centering
    \includegraphics[width=\textwidth]{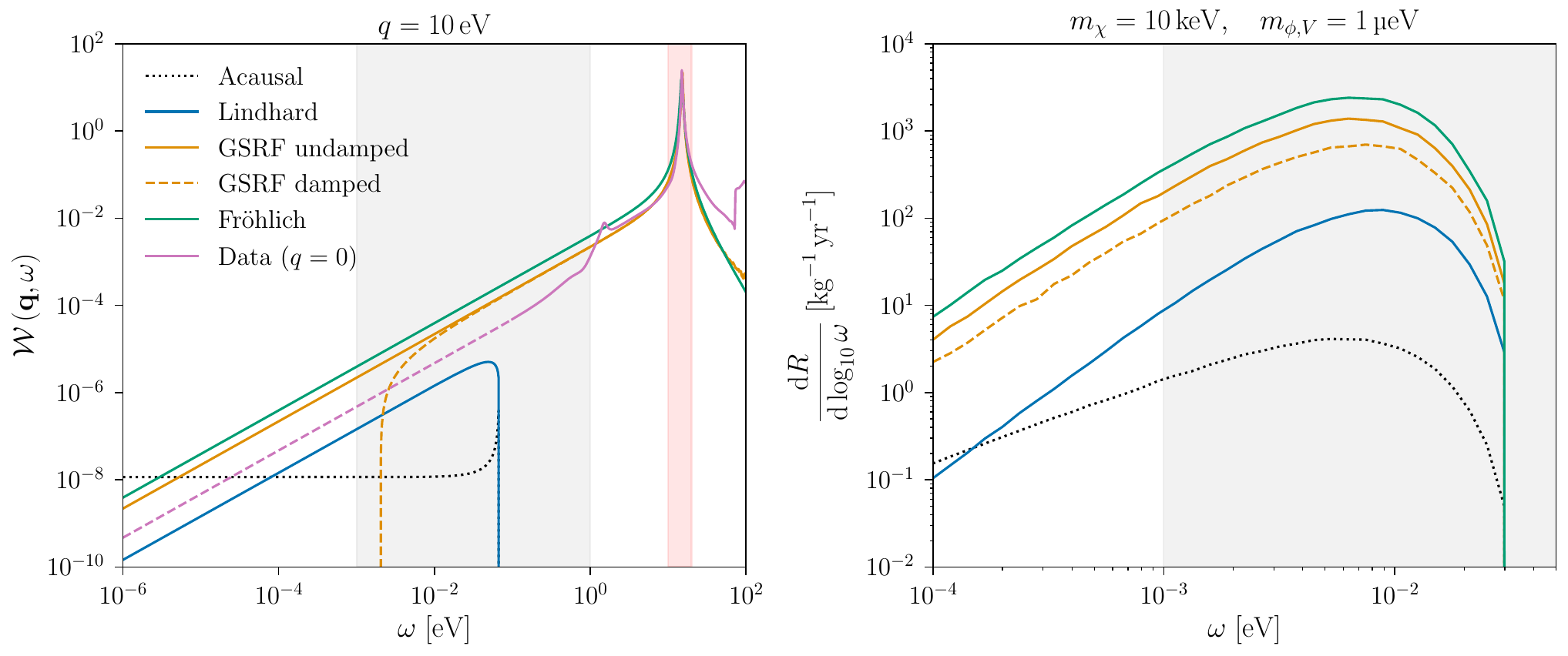}
    \caption{\textbf{Left:} loss function for each of several models for Al, for $q = \SI{10}{\eV}$. The curve labeled `Acausal' shows the loss function used in \cite{Hochberg:2015fth}, which involves an unphysical choice of branch cut in the complex logarithm. The Fr\"ohlich model fit is the same as that shown in \cref{fig:measuredloss}, for which measured data are only available within the red band. The Lindhard model is the RPA dielectric function \cref{eq:e-rpa-fw} with $\Gamma_p = 0$, and the GSRF models use fit parameters for \cref{eq:eps-gsrf} from Ref.~\cite{gibbons1976line}, with `undamped' corresponding to $\Im G = 0$ and `damped' corresponding to $\Im G \neq 0$. The damped curve becomes negative at small $\omega$, which is an unphysical consequence of the GSRF model. The curve labeled `Data' shows the fit to $q=0$ measurements provided by Ref.~\cite{sun2016calculations}. We use dashes to indicate the continuation of the fit beyond the range of measured data. The gray band shows the reference range of \SI{1}{\meV}--\SI{1}{\eV} deposits. \textbf{Right:} recoil spectra corresponding to each of these loss functions, assuming $(m_\chi,\,m_{\phi,V})=(\SI{10}{\keV},\,\SI{1}{\micro\eV})$ and $\overline\sigma_e=\SI{e-39}{\centi\meter^2}$.
    }
    \label{fig:superconductor-losses-spectra}
\end{figure*}

Measurements of the loss function in the vicinity of the plasmon peak are available in the literature for certain materials, so it is already possible to fit the Fr\"ohlich model directly to data and to assess the significance of the plasmon tail at $\omega\ll\omega_p$. \Cref{fig:measuredloss} (left) shows such a fit to measurements in Al. While the fit is excellent in the vicinity of the plasmon peak, the behavior at $\omega\ll\omega_p$ should be viewed only as a benchmark: other physical effects contribute at these energies, notably those encapsulated by the Lindhard dielectric function which incorporates electron screening effects. See \cref{fig:superconductor-losses-spectra} and \cref{sec:s-reach-projections} below for further details.

High-precision measurements of the loss function at nonzero $q$ have also been performed for Si using X-ray scattering~\cite{weissker2010dynamic}. The plasmon is clearly visible at small $q$, but here we focus on the behavior at large $q$. \Cref{fig:measuredloss} (center) shows the measured loss function along the [100] crystal direction (solid lines), compared to the RPA loss function for the homogeneous electron gas taking $\omega_p = \SI{16.67}{\eV}$ for the measured plasmon frequency \cite{kundmann1988study}. While semiconductors and insulators do not, strictly speaking, have a Fermi velocity at zero temperature where there are no free carriers, we may regard $v_F$ as a tuneable parameter which governs the behavior of the loss function at small $\omega$. With $v_F = 8.6 \times 10^{-3}$, on the same scale as $v_F$ for typical metals, the fit is quite good, especially for $\omega < \SI{25}{\eV}$. On the other hand, at $q = \SI{10}{\keV}$, the RPA loss function vanishes identically for $\omega < \SI{12}{\eV}$, which is likely unphysical given that atomic tight-binding wavefunctions have support in this kinematic range. The purpose of this comparison is not to advocate for using this extremely simplified model---indeed, data should be used to compute DM rates whenever possible---but rather to demonstrate how in the absence of data a simple model may provide an accurate estimate for the light-mediator spectrum for $\omega \in [\SI{5}{\eV}, \SI{10}{\eV}]$, where the rate integral is dominated by $q \in [\SI{5}{\keV}, \SI{10}{\keV}]$. Indeed, the success of the RPA model suggests that this part of the spectrum from scattering in any semiconductor or insulator with eV-scale bandgaps is nearly universal, determined only by the valence electron density and an effective Fermi velocity. This model may be seen as an extension, accounting for screening, of earlier simplified models for scattering in semiconductors using atomic orbitals or tight-binding wavefunctions \cite{Graham:2012su,Lee:2015qva}.

To complete our survey of sample loss functions, we show in Fig.~\ref{fig:measuredloss} (right) the measured loss function at $q = 0$ for \URuSi\ along the $a$ and $c$ crystal axes, measured with Fourier transform infrared spectrometry \cite{bachar2016detailed}. \URuSi\ has been extensively studied for decades~\cite{Mydosh_2014} due to its unusual `hidden order' below \SI{17.5}{\kelvin}, and thus has been synthesized as ultra-pure single crystals. Below $T_c = \SI{1.5}{\kelvin}$ it behaves as a conventional superconductor~\cite{palstra1985superconducting}. A number of features are present below \SI{20}{\meV} which may be interpreted as heavy-fermion plasmons, as we discuss in the main text. Based on this interpretation, to perform our rate estimates in the main text, we extrapolate the loss function as independent of $q$ out to $q = q_c \simeq \SI{100}{\eV}$. Indeed, this is the standard approximation made in scattering experiments near the plasmon pole \cite{kundmann1988study}. Then, we see from~\cref{eq:spectrum} that the spectrum is largely determined by the shape of the zero-momentum loss function $\mathcal{W}(\omega)$, with the inverse mean speed $\eta$ only serving to enforce the kinematic condition $q > \omega/v_\chi$. All of the approximations we have made may easily be dropped once momentum-resolved data on $\mathcal{W}(\bb q, \omega)$ within the DM regions shown in Fig.~\ref{fig:kinematics} is available.

It is also interesting to note that at larger $\omega$, the loss function is linear to an excellent approximation, in the $c$ direction above \SI{20}{\meV} and in the $a$ direction above \SI{50}{\meV}. In \cref{fig:measuredloss} we show a linear fit to both loss functions with zero offset. In the Fr\"ohlich model \cref{eq:frolich}, the plasmon tail gives a loss function $\mathcal{W}_F(q = 0, \omega) \approx \omega \times (\Gamma_p/\omega_p^2)$ at small $\omega$. The slope of the linear fit is consistent with $\Gamma_p/\omega_p \simeq 0.1-0.2$ and $\omega_p \simeq \SI{15}{\eV}$, which would be reasonable parameters for the ordinary valence electron plasmon in a generic metal. This data therefore provides some preliminary indication that the linear tail of the plasmon in ordinary superconductors like Al may extend down to the meV scale. We emphasize again that dedicated measurements are needed to confirm this. 

\ssubsection{Semiconductor spectrum in the free-electron gas approximation}

\begin{figure}
    \centering
    \includegraphics[width=0.45\textwidth]{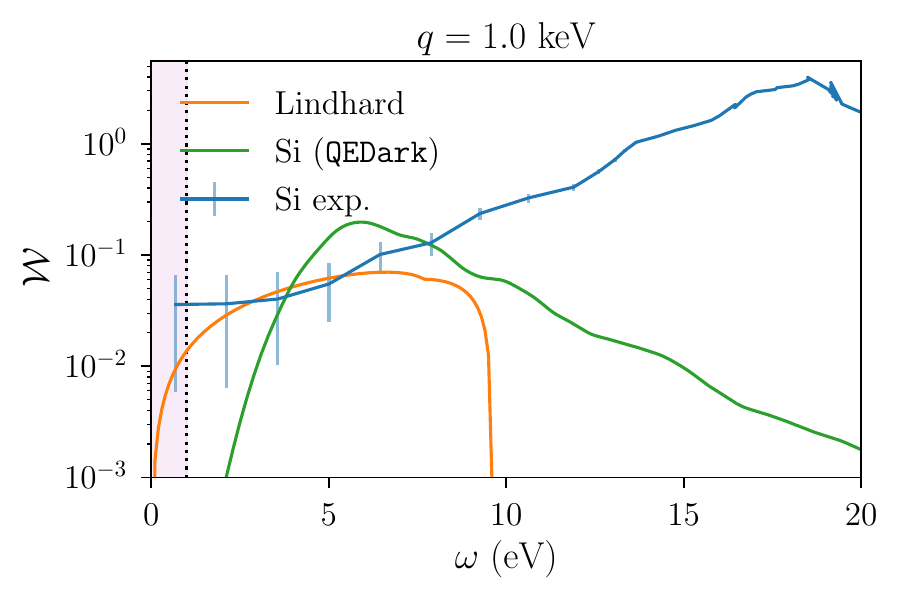}
    \includegraphics[width=0.45\textwidth]{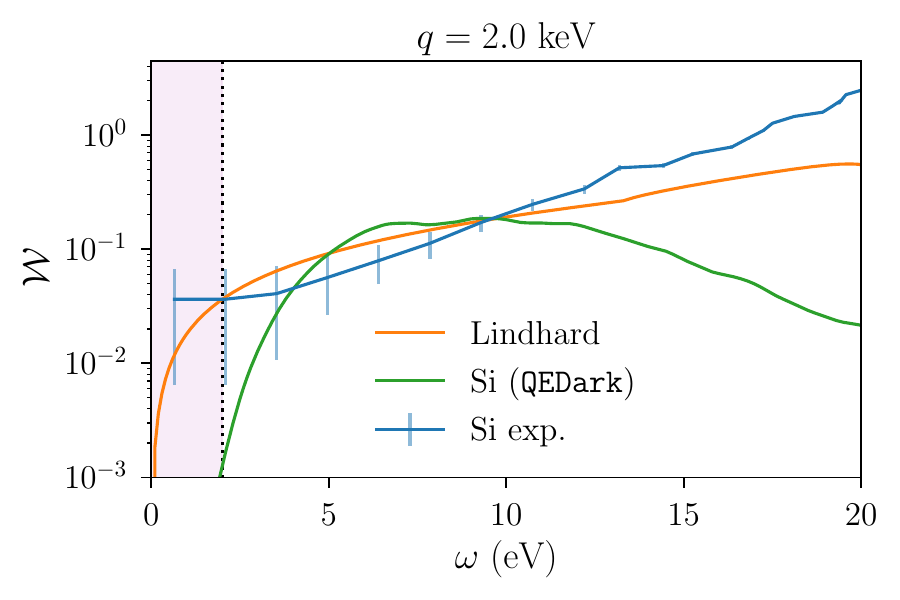}
    \includegraphics[width=0.45\textwidth]{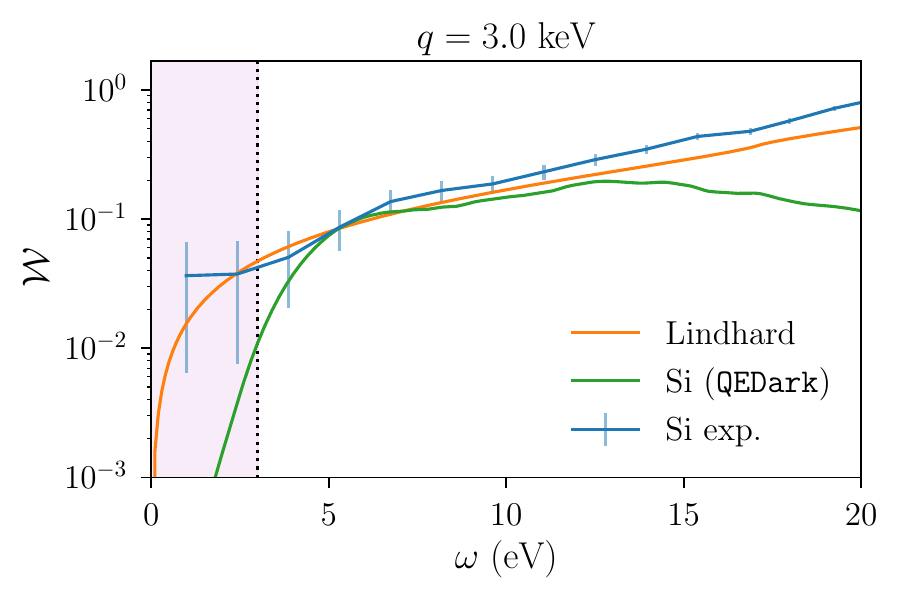}
    \includegraphics[width=0.45\textwidth]{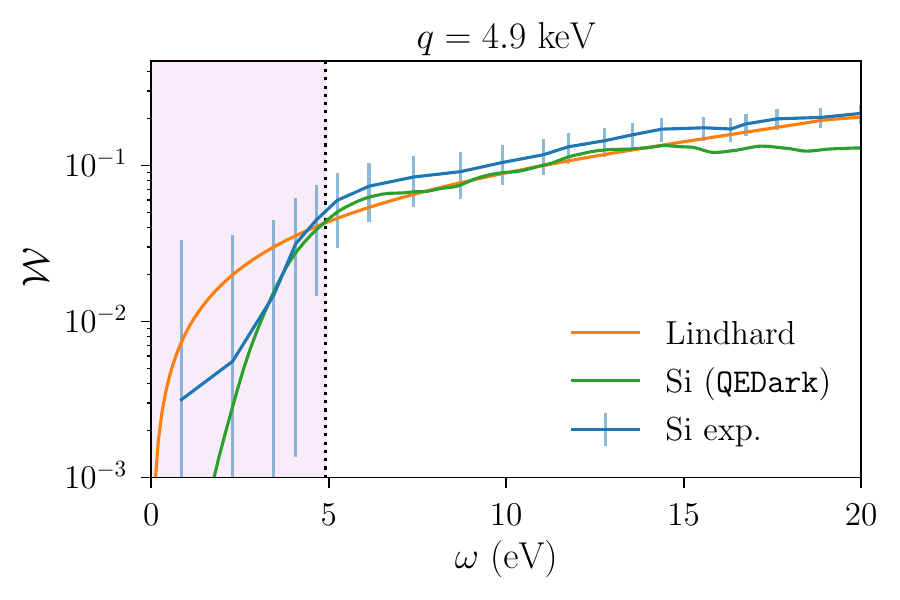}
    \includegraphics[width=0.45\textwidth]{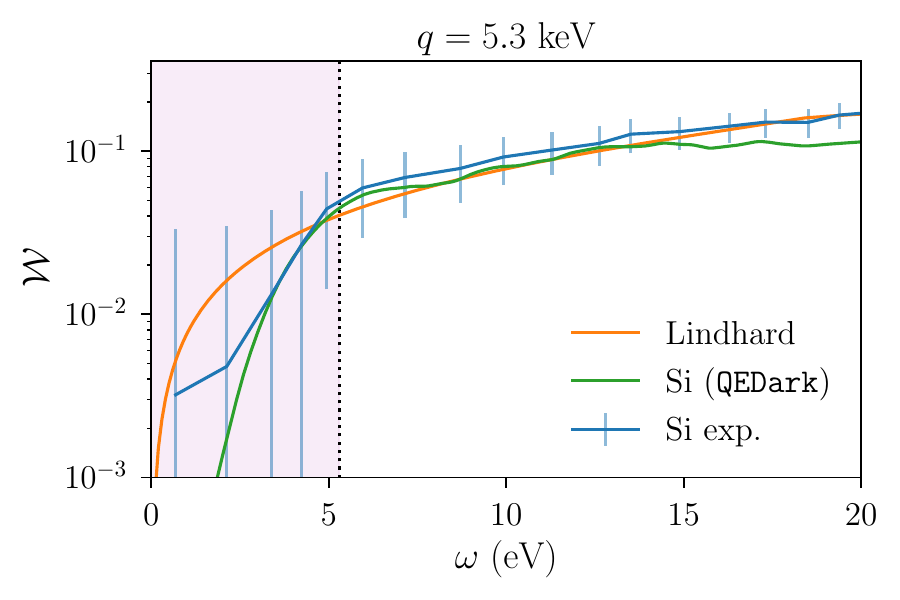}
    \includegraphics[width=0.45\textwidth]{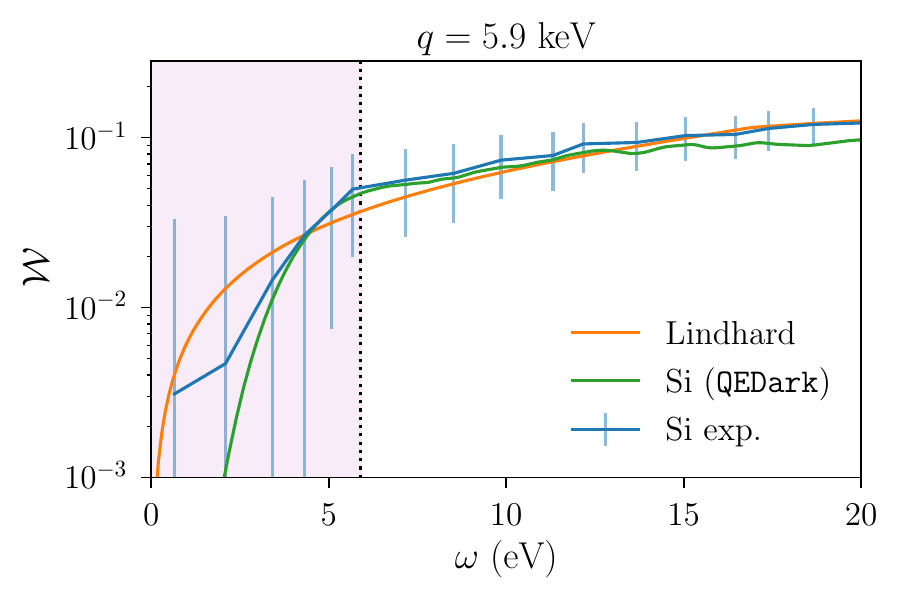}
                    
    \caption{Loss function comparisons in Si for various $q$, as a function of $\omega$. Error bars indicate the accuracy with which the data points could be transcribed from Ref.~\cite{weissker2010dynamic}. The shaded purple region represents the kinematically-allowed region for $v_\chi = 10^{-3}$. The measured loss function agrees fairly well with both the loss function computed from the single-particle basis from \texttt{QEdark}~\cite{Essig:2015cda} and the Lindhard FEG approximation in the range 5--\SI{10}{\eV} for $\omega$, but there are large differences at both small $\omega$ near the gap, and near the plasmon energy $\omega_p \simeq \SI{17}{\eV}$ for small $q$.}
    \label{fig:QEDark-loss-function}
\end{figure}

\begin{figure}
    \includegraphics[width=\textwidth]{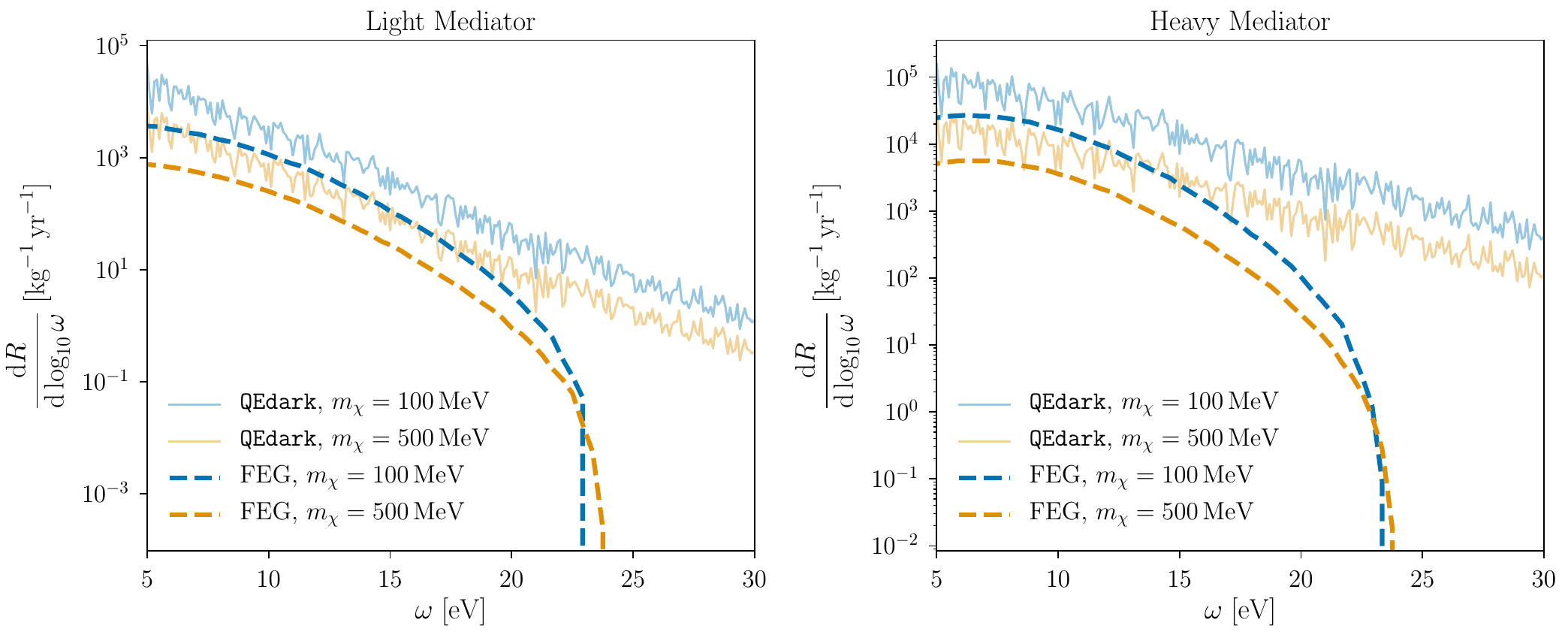}
    \caption{Recoil spectra in Si at fixed $\overline\sigma_e = \SI{e-37}{\centi\meter^2}$, for light and heavy mediators (scalar or vector). Solid curves are computed with \texttt{QEdark} \cite{Essig:2015cda}. Dashed curves are computed from \cref{eq:THEANSWER} with the Lindhard RPA loss function, \cref{eq:e-rpa-fw} with $v_F = 8.6 \times 10^{-3}$, $k_F = m_e v_F$, and $\omega_p = \SI{16.67}{\eV}$.}
    \label{fig:silicon-spectrum}
\end{figure}

In order to relate the energy loss function to the crystal form factor \cite{Essig:2015cda,Trickle:2019nya}, we compare

\begin{equation}
    \Gamma(\bb v_\chi) = \int\frac{\du^3\bb q}{(2\pi)^3} \left|V(q)\right|^2
        \mathcal{S}(\bb q, \omega)
\end{equation}
 to \cref{eq:THEANSWER}, which gives the relation between the dynamic structure factor $\mathcal{S}(\bb q, \omega)$ defined in Ref.~\cite{Trickle:2019nya} and the loss function via

\begin{equation}
    \mathcal{S}(\bb q,\omega) = \frac{2q^2}{e^2}\mathcal{W}(\bb q,\omega).
    \label{eq:SWdef}
\end{equation}

On the other hand, the dynamic structure factor in a semiconductor, computed in the basis of single-particle states, can be related to the crystal form factors $|f_{ii'\bb k \bb k' \bb G}|^2$ via \cite{Essig:2015cda,Trickle:2019nya}

\begin{equation}
    \mathcal{S}(\bb q,\omega) = 2 \sum_{i,i',\bb G} \int_{\mathrm{BZ}} \frac{d^3\bb k}{(2\pi)^3}\frac{d^3 \bb k'}{(2\pi)^3} 2\pi\delta(E_{i'\bb k'}-E_{i\bb k} - \omega) 2\pi \delta(|\bb k'-\bb k + \bb G|-q) |f_{ii'\bb k \bb k' \bb G}|^2,
\end{equation}
where the momentum integral is taken over the first BZ, $\mathbf{G}$ runs over all reciprocal lattice vectors, and $i$ and $i'$ run over all valence and conduction bands, respectively. Thus we can compute an equivalent loss function from \texttt{QEdark}~\cite{Essig:2015cda} crystal form factors by

\begin{equation}
    \mathcal{W}(\bb q,\omega) = \frac{e^2}{q^2} \sum_{i,i',\bb G} \int_{\mathrm{BZ}} \frac{d^3\bb k}{(2\pi)^3}\frac{d^3 \bb k'}{(2\pi)^3} 2\pi\delta(E_{i'\bb k'}-E_{i\bb k} - \omega) 2\pi \delta(|\bb k'-\bb k + \bb G|-q) |f_{ii'\bb k \bb k' \bb G}|^2
    \label{eq:WQEDark}
\end{equation}

Using \cref{eq:WQEDark}, we can compare the measured loss function to the loss function computed in the single-particle basis by \texttt{QEdark}, as well as the Lindhard dielectric function for the FEG with the best-fit $v_F$ in \cref{fig:measuredloss}. The results are shown in \cref{fig:QEDark-loss-function}. Note that for a given $q$, the range of $\omega$ which is accessible is $\omega < q v_\chi$, which only comprises a small piece of the total support of $\mathcal{W}(\bb q, \omega)$. Regardless, we see that \texttt{QEdark} tends to slightly underpredict the measured loss in the kinematically-allowed region. Furthermore, \texttt{QEdark} accurately reproduces the measured loss in the near-gap region $\omega \in [\SI{1}{eV}, \SI{5}{\eV}]$ where Lindhard fails to do so, as expected. On the other hand, \texttt{QEdark} fails to capture the plasmon which is seen in the measured loss function because the single-particle band structure states do not account for collective effects.

Overall, though, the nearly-linear shape of the measured loss function in the range $\omega \in [\SI{5}{\eV}, \SI{15}{\eV}]$ is reproduced fairly well by the Lindhard model, and matches that of \texttt{QEdark}. We therefore expect that the spectral shape (though perhaps not the normalization) will be captured in this energy range by the simple Lindhard model for the loss function. Moreover, since the Lindhard model loss function goes to zero at sufficiently large $q$ for small $\omega$, and since the rate recieves contributions from \emph{all} $q > \omega/v_\chi$, we expect the Lindhard approximation to be best for a light mediator which weights the rate integrand by $|V(q)|^2 \propto 1/q^4$. The results are shown in \cref{fig:silicon-spectrum}. Indeed, the Lindhard FEG model matches the spectrum fairly well for the light mediator, roughly independent of the DM mass as long as the DM kinetic energy is well above the gap. The spectrum for a heavy mediator is a poorer match, especially at large $\omega$ where the kinematic mismatch between the FEG and the bound atomic wavefunctions becomes more important. We emphasize once again that these simple arguments are \emph{not} meant to replace a measurement of $\mathcal{W}$ in the relevant kinematic range, which would predict the spectrum unambiguously. However, they do highlight a qualitative understanding of the spectrum in a limited energy range based on simple material properties like the effective $v_F$, which may be useful for identifying other detector materials suitable for DM-electron scattering. Furthermore, the part of the spectrum where the FEG model performs best corresponds to the 2-electron bin in Si, which is of considerable practical importance to experiments: the 1-electron bin is typically dominated by backgrounds such as leakage current and Cherenkov radiation \cite{Du:2020ldo}, while the rates in the bins with 3 or more electrons drop precipitously, at least based on estimates from the single-particle loss function. Integrating the FEG spectra from a threshold of $\omega = 4.7 \ {\rm eV}$, corresponding to a $2e^{-}$ threshold in the model of Ref.~\cite{Essig:2015cda}, we obtain the reach curve shown in \cref{fig:reach} in the main text.

\ssection{Updated reach projections for superconductors}
\label{sec:s-reach-projections}

In Ref.~\cite{Hochberg:2015pha}, the scattering rate in a superconductor is first computed treating the electrons as free particles, with screening included afterwards in Ref.~\cite{Hochberg:2015fth} via a correction to the matrix element. We now show that the result of Ref.~\cite{Hochberg:2015fth} at $T=0$ is exactly reproduced by our \cref{eq:THEANSWER} when $\epsilon(\bb q,\omega)$ is taken to be the Lindhard dielectric function in the limit of vanishing plasmon width.

In a relativistic formalism for single-particle scattering, the superconductor scattering rate is given by 
\begin{equation}
    \label{eq:traditional-rate}
    \Gamma(\bb v_\chi) = \int\frac{\du^3\bb p_\chi^\prime}{(2\pi)^3}
        \frac{\bigl\langle\left|\mathcal M\right|^2\bigr\rangle}
            {16E_\chi E_\chi^\prime E_e E_e^\prime}
        \frac{S(\bb q, \omega)}{\left|\epsilon(\bb q, \omega)\right|^2},
\end{equation}
where $\bb q\equiv \bb p_\chi - \bb p_\chi^\prime$ denotes the 3-momentum transfer, $\bb p_\chi^\prime$ denotes the momentum of the scattered dark matter particle in the final state, and $S(\bb q,\omega)$ (not to be confused with the dynamic structure factor defined in \cref{eq:SWdef} above) characterizes the available phase space, to be defined shortly. The presence of $|\epsilon|^2$ in the denominator of \cref{eq:traditional-rate} accounts for screening and was treated in Ref.~\cite{Hochberg:2015fth} as an in-medium modification to the dark photon propagator. In the non-relativistic limit, any interaction of the class considered in \cref{eq:potentialSM} gives rise to a matrix element of the form
\begin{equation}
    \frac{\bigl\langle\left|\mathcal M\right|^2\bigr\rangle}
         {16E_\chi E_\chi^\prime E_e E_e^\prime} \simeq 
        \left(\frac{g_\chi g_e}{q^2 + m_{\phi,V}^2}\right)^2
    = \left|V(q)\right|^2
    ,
\end{equation}
where $q=|\bb q|$. \Cref{eq:traditional-rate} is trivially transformed to an integral over $\bb q$, and the rate becomes
\begin{equation}
    \Gamma(\bb v_\chi) = \int\frac{\du^3\bb q}{(2\pi)^3} \left|V(q)\right|^2
        \frac{S(\bb q, \omega)}{\left|\epsilon(\bb q, \omega)\right|^2}
    .
\end{equation}
Thus, to agree with \cref{eq:THEANSWER}, it is sufficient to have
\begin{equation}
    \label{eq:s-condition}
    S(\bb q, \omega) = \frac{2q^2}{e^2}\Im\epsilon(\bb q, \omega)
    .
\end{equation}

\Cref{eq:s-condition} holds exactly in the low-temperature limit for the form of $S$ used in Refs.~\cite{Hochberg:2015pha,Hochberg:2015fth}, where the superconductor is treated as a free electron gas. In this case, $S$ is given by
\begin{equation}
    \label{eq:s-definition}
    S(\bb q,\omega) = 2\int
    \frac{\du^3\bb p_e}{(2\pi)^3}\frac{\du^3\bb p_e^\prime}{(2\pi)^3}
    (2\pi)^4\delta^4(P_\chi + P_e - P_\chi^\prime - P_e^\prime)
    f_{\mathrm{FD}}(E_e)\left[1 - f_{\mathrm{FD}}(E_e^\prime)\right]
    ,
\end{equation}
where $f_{\mathrm{FD}}$ is the Fermi--Dirac distribution and the $P_i$ denote 4-momenta. We reserve $p_i$ for the magnitudes of 3-momenta. The integration over $\bb p_e^\prime$ is readily performed using the 3-momentum delta function. Writing the $\bb p_e$ integral in spherical coordinates and performing the trivial integral over the azimuthal angle produces
\begin{equation}
    S(\bb q, \omega) = 2\int
        \frac{p_e^2\dd p_e\dd(\cos\theta)}{(2\pi)^2}
        \delta\left(
            \omega - \frac{q^2 + 2p_eq\cos\theta}{2m_e}
        \right)
        f_{\mathrm{FD}}(E_e)\left[1 - f_{\mathrm{FD}}(E_e^\prime)\right]
    ,
\end{equation}
where $\theta$ denotes the angle between $\bb p_e$ and $\bb q$. The remaining delta function can be used to evaluate the integral over $\cos\theta$, but here care must be taken to enforce $|\cos\theta|\leq 1$. With the appropriate Heaviside function, the final integral becomes
\begin{equation}
    S(\bb q, \omega) = \int\du p_e\,\frac{m_ep_e}{\pi q}
        \left[1 - f_{\mathrm{FD}}(E_e^\prime)\right]
        \Theta\left(
            1 - \left|
                \frac{2m_e\omega - q^2}{2p_eq}
            \right|
        \right)
    .
\end{equation}
Now the zero-temperature Fermi--Dirac distribution can be inserted and the integral can be performed analytically. The result is
\begin{equation}
    \label{eq:s-general}
    S(\bb q, \omega) = \frac{m_e^2}{\pi q}
    \begin{cases}
        \omega
            & 0 < \omega < \left|E_-\right| \\
        E_F - \frac{(E_q-\omega)^2}{4E_q}
            & \left|E_-\right| < \omega
                < E_+ \\
        0 & \mathrm{otherwise},
    \end{cases}
\end{equation}
where $E_q\equiv q^2/2m_e$ and $E_\pm = E_q \pm qv_F$. The conditions in \cref{eq:s-general} are equivalent to those in \cref{eq:im-e-rpa}, {\it i.e.}, the imaginary part of the Lindhard dielectric function in the limit that the plasmon is infinitely long-lived. \Cref{eq:s-condition} follows by direct comparison.

Given this agreement between the single-particle and dielectric-function formalisms, \cref{eq:THEANSWER} can reproduce prior calculations of the scattering rate in superconductors; essentially, the final-state phase space integral is pre-computed in $\Im(\epsilon)$. However, \cref{eq:THEANSWER} is more flexible than the traditional calculation in that we are not limited to the narrow-plasmon limit of the Lindhard dielectric function. Any model or measurement of the loss function can be inserted directly in \cref{eq:THEANSWER}.

To evaluate the event rate in a superconducting detector, we take the velocity of the DM in the galactic frame to have a modified Maxwell--Boltzmann distribution,
\begin{equation}
    f(\bb v_\chi) \propto \exp\left(-\bb v_\chi^2/v_0^2\right)
        \Theta\bigl(v_{\mathrm{esc}} - |\bb v_\chi|\bigr)
    .
\end{equation}
For our reach projections, we take $v_0 = \SI{220}{\kilo\meter/\second}$ and $v_{\mathrm{esc}} = \SI{550}{\kilo\meter/\second}$, and we take Earth to have a velocity $v_E = \SI{232}{\kilo\meter/\second}$ in the galactic frame. This matches the conventions of Ref.~\cite{Hochberg:2017wce}. In order to facilitate comparison with other results in the literature, we also show some results with $v_E = 0$ and $v_{\mathrm{esc}} = \SI{500}{\kilo\meter/\second}$, matching the conventions of {\it e.g.} Refs.~\cite{Hochberg:2015pha,Hochberg:2015fth}. We refer to this as the `simple halo' scenario. Finally, for illustrative purposes, we show selected results for a hypothetical halo with $v_0=v_{\mathrm{esc}}=\SI{e4} {\kilo\meter/\second}$ and $v_E=0$. In this `fast DM' scenario, the plasmon peak is kinematically accessible, and this is directly visible as a feature in the recoil spectrum.

\begin{figure*}
    \centering
    \includegraphics[width=\textwidth]{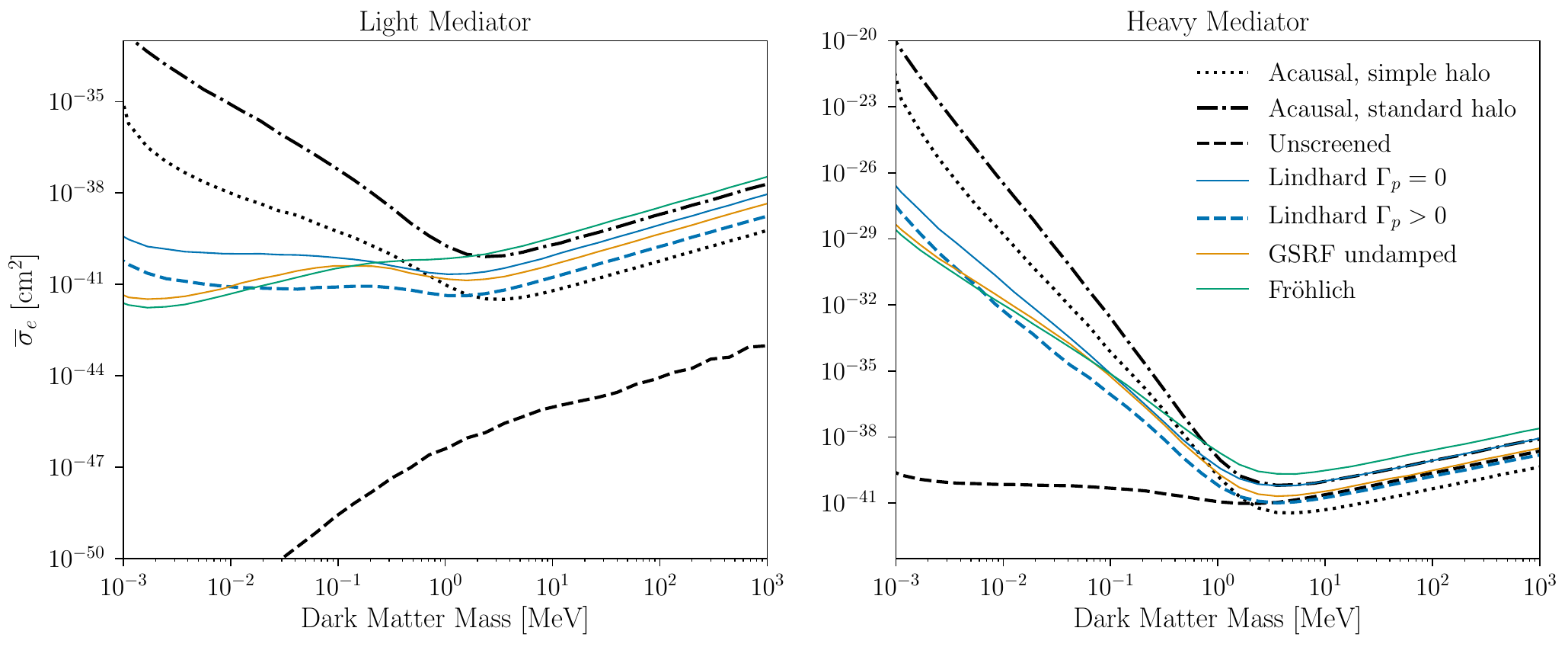}
    \caption{Projected reach for an aluminum superconductor target for several forms of the loss function for scalar or vector mediators. The dotted curve is computed with the simple halo model ($v_E=0$, $v_{\mathrm{esc}} = \SI{500}{\kilo\meter/\second}$), and all others assume the standard halo model. The dashed line (`unscreened') is computed in the single-particle formalism with no correction for screening, {\it i.e.}, without the factor of $|\epsilon|^2$ in the denominator of \cref{eq:traditional-rate}; this is unphysical for any spin-independent DM-electron interaction.}
    \label{fig:superconductor-reach}
\end{figure*}

\begin{figure*}
    \centering
    \includegraphics[width=\textwidth]{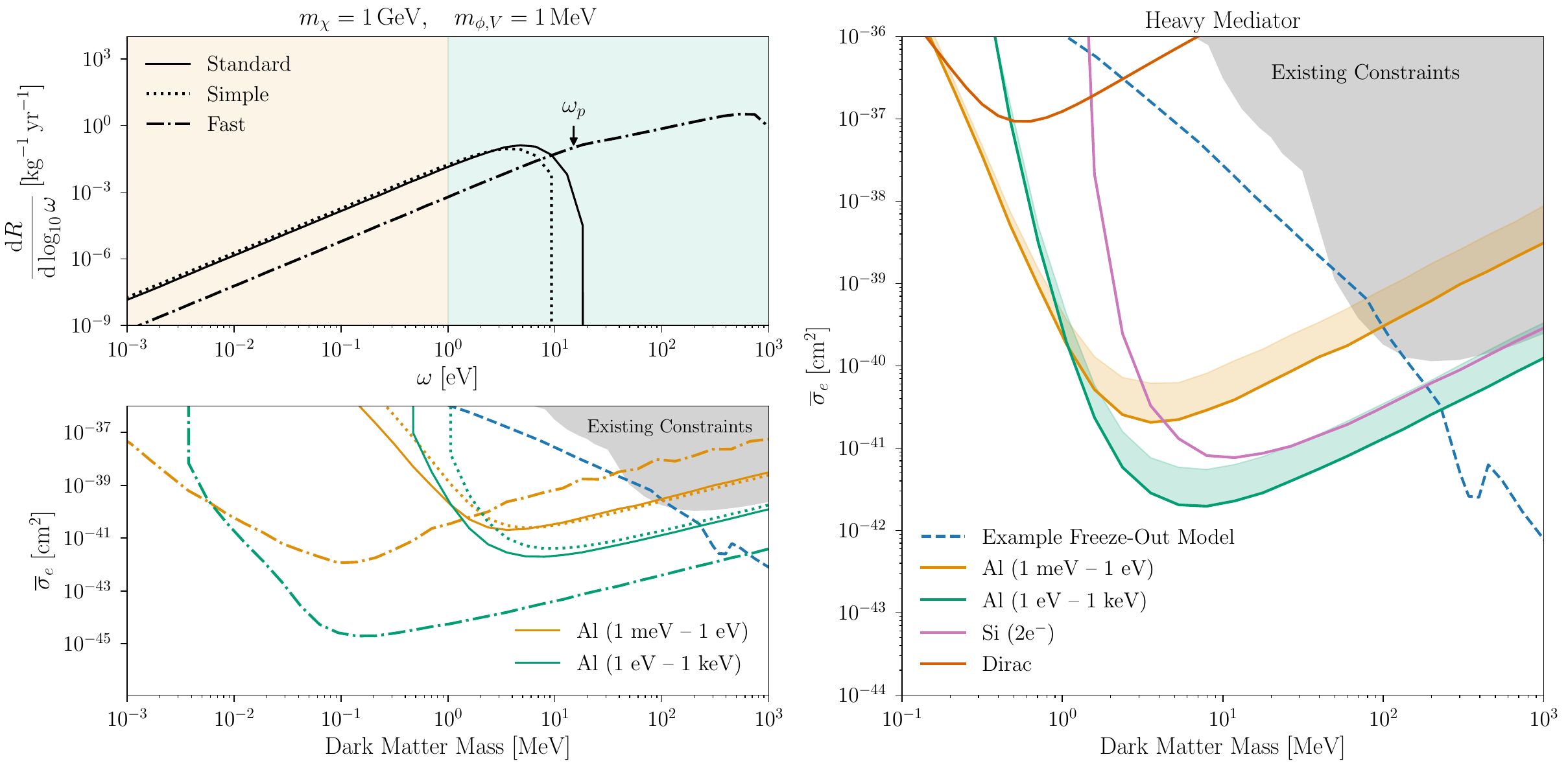}
    \caption{\textbf{Top left:} recoil spectra in an Al superconductor for $(m_\chi,\,m_\phi)=(\SI{1}{\giga\eV},\,\SI{1}{\mega\eV})$ and $\overline\sigma_e=\SI{e-42}{\centi\meter^2}$ assuming the GSRF loss function without damping for several DM velocity distributions. See text for details. The fast halo scenario is unrealistic and is shown for illustrative purposes only: in this case, the plasmon peak is kinematically accessible, and the recoil spectrum exhibits a corresponding kink at $\omega = \omega_p$. The shaded areas indicate two fiducial experimental configurations, one sensitive to deposits \SI{1}{\meV}--\SI{1}{\eV} (orange), and the other sensitive to deposits \SI{1}{\eV}--\SI{1}{\keV} (green). \textbf{Bottom left:} projected reach in an Al superconductor for a 1~kg-yr exposure assuming a heavy mediator. Orange curves show the reach for the low-threshold scenario, and green curves show the reach for the high-threshold scenario. Projections for the standard halo, simple halo, and fast halo scenarios are shown by the solid, dotted, and dot-dashed curves, respectively. \textbf{Right:} projected reach for a 1~kg-yr exposure of a Dirac material, Si, and the two Al superconductor configurations assuming a heavy scalar or vector mediator. The parameters of the Dirac material are taken as in \cref{fig:reach}, with gap $2\Delta = \SI{20}{\meV}$, Fermi velocity $v_F = 4\times10^{-4}$, background dielectric constant $\kappa = 40$, and Dirac band cutoff $\omega_{\mathrm{max}} = \SI{0.5}{\eV}$. The projected reach for Si assumes a two-electron ionization threshold. The projected reach of \URuSi~lies above the top edge of the plot. All curves assume the standard halo model. For the Al target, the shaded regions indicate the range of variation in different models of the loss function. The solid lines are computed using the GSRF loss function without damping, and the top of each shaded band is computed using the Lindhard loss function. An example of the target parameter space for thermal freeze-out through a heavy dark photon mediator \cite{Battaglieri:2017aum} is shown in dashed blue.}
    \label{fig:heavy-reach}
\end{figure*}

The various models for the loss functions in Al are shown in \cref{fig:superconductor-losses-spectra}, together with the corresponding DM recoil spectra for a kg-yr exposure. The undamped GSRF model and the Lindhard model with $\Gamma_p=0$ correspond to the boundaries of the shaded region in \cref{fig:reach}. We also show the result obtained by choosing the acausal branch in the Lindhard dielectric function. It is clear from \cref{fig:superconductor-losses-spectra} that at low energies, a naive extrapolation of the plasmon tail dominates over the Lindhard loss function with its infinitely long-lived plasmon. Moreover, the energy range of interest for light DM detection is precisely where the effects of damping in the GSRF loss function become important. While the loss functions given here are valuable benchmarks, the true loss function likely falls somewhere between the Lindhard result and the plasmon tail. This is also suggested by fitting the measurements of Ref.~\cite{sun2016calculations}, which go down to $\omega = \SI{100}{\milli\eV}$ and lie somewhat below the plasmon tail. (See the purple line in \cref{fig:superconductor-losses-spectra}.) To accurately predict the DM scattering rate, it is both essential and feasible to \emph{measure} the loss function in the entire relevant regime of $\SI{1}{\milli\eV}<\omega<\SI{1}{\eV}$.

\Cref{fig:superconductor-reach} shows updated reach curves for an aluminum superconductor target alongside the results of Refs.~\cite{Hochberg:2015pha,Hochberg:2015fth}. The reach curves are specified with respect to a reference cross section defined by
\begin{equation}
    \overline\sigma_{e} =
        \frac{16\pi\mu_{e\chi}^2\alpha_{e} \alpha_\chi}
             {\left((\alpha_{\mathrm{EM}} m_e)^2 + m_\phi^2\right)^2},
\end{equation}
where $\mu_{\chi e}$ denotes the reduced mass of the electron--DM system, $m_\phi$ is the mediator mass, and $\alpha_{e,\chi} = g_{e,\chi}^2/(4\pi)$ in terms of the couplings which define the potential in \cref{eq:potentialSM}. In \cref{fig:superconductor-reach}, `light mediator' means $m_\phi \ll \alpha_{\mathrm{EM}} m_e$ (defined with respect to the ordinary electromagnetic fine-structure constant $\alpha_{\mathrm{EM}} \simeq 1/137$) and `heavy mediator' means $m_\phi \gg \alpha_{\mathrm{EM}} m_e$. All reach projections are computed in the zero-temperature limit and assume that the detector is sensitive to deposits between \SI{1}{\meV} and \SI{1}{\eV}. We show reach curves for a high-threshold experiment sensitive to deposits \SI{1}{\eV}--\SI{1}{\keV} for a heavy mediator in \cref{fig:heavy-reach}, along with recoil spectra for selected model points. We also illustrate the appearance of a feature in the recoil spectrum at $\omega_p$ in the fast halo model, where the plasmon peak is kinematically accessible. To facilitate comparison with the literature, we show reach curves corresponding to an event rate of \SI{3}{\per\kilogram\per\year}, which corresponds roughly to a 95\% C.L. constraint.

\Cref{fig:superconductor-reach} in particular underscores the importance of properly treating the material response. For any interaction of the kind we consider in this work, screening is significant at low DM mass or for a light mediator. However, the implementation of screening in Ref.~\cite{Hochberg:2015fth} overestimated the size of the effect for a vector mediator: at the lowest DM masses, the causal branch choice in the logarithms of \cref{eq:e-rpa-fw} yields a rate as much as seven orders of magnitude greater than that produced by the acausal choice. Furthermore, accounting for the non-zero width of the plasmon peak further enhances the rate by an order of magnitude or more. The lingering uncertainty in analytical predictions of the loss function can be easily resolved by directly measuring the loss function in promising target materials.

\end{document}